\documentclass[
aps,%
11pt,%
final,%
notitlepage,%
oneside,%
twocolumn,%
nobibnotes,%
nofootinbib,%
superscriptaddress,%
noshowpacs,%
centertags,%
landscape]%
{revtex4}

\newcommand{\HI}{HI}

\newcounter{qub}
\newcommand{\qq}{\addtocounter{qub}{1}\arabic{qub}}
\usepackage{lscape}
\usepackage{amsmath}
\usepackage{eufrak}
\newcommand{\M}{\mathfrak{M}}

\begin{document}

\title{Study of Galaxies in the  Lynx--Cancer Void. IV.~Photometric Properties
}

\author{\firstname{Yu.~A.}~\surname{Perepelitsyna}}
\email[E-mail:]{jlyamina@yandex.ru} 
\affiliation{Special Astrophysical Observatory, 
Russian Academy of Sciences, Nizhnii Arkhyz, 369167 Russia}

\author{\firstname{S.~A.}~\surname{Pustilnik}}
\email[E-mail:]{sap@sao.ru}
\affiliation{Special Astrophysical Observatory, 
Russian Academy of Sciences, Nizhnii Arkhyz, 369167 Russia}

\author{\firstname{A.~Yu.}~\surname{Kniazev}}
\email[E-mail:]{akniazev@saao.ac.za}
\affiliation{South African Astronomical Observatory, Cape Town, 7935 
South Africa}\affiliation{Southern African Large Telescope, Cape 
Town, 7935 South Africa}\affiliation{Lomonosov Moscow State University, 
Moscow, 119991 Russia}

\received{December 13, 2013}  \revised{May 14, 2014}


\begin{abstract}
We present the results of a photometric study of 85 objects from
the updated sample of galaxies residing in the nearby Lynx-Cancer
void. We perform our photometry on  $u, g, r,$ and $i$-band images
of the Sloan Digital Sky Survey. We determine model-independent
galaxy parameters such as the integrated magnitudes and colors,
effective radii and the corresponding  surface brightness values,
optical radii and Holmberg radii. We analyze the radial surface
brightness profiles to determine the central brightness values and
scale lengths of the model disks. We analyze the colors of the
outer parts of the galaxies and compare them with model
evolutionary tracks computed using the {\tt PEGASE\,2} software
package. This allowed us to estimate the time  $T_{\rm SF}$
elapsed since the onset of star formation, which turned out to be
on the order of the cosmological time $T_{\mathrm 0}$ for the
overwhelming majority of the galaxies studied. However, for
13~galaxies of the sample the time $T_{\rm SF}$ does not exceed
$T_{\mathrm 0}/2 \sim7$~Gyr, and for 7 of them \mbox{$T_{\rm SF}
\lesssim3.5$~Gyr}. The latter are mostly unevolved objects
dominated by low-luminosity galaxies with  $M_{B} > -13.2$. We use
the integrated magnitudes and colors to estimate the stellar
masses of the galaxies. We estimate the parameter $\M({\rm
H\,I})/L_{B}$ and the gas mass fractions for void galaxies with
known H\,I-line fluxes. A small subgroup (about 10\%) of the
gas-richest void galaxies   with $\M({\rm H\,I})/L_{B} \gtrsim
2.5$ has gas mass fractions that reach 94-99\%. The
outer regions of many of these galaxies show atypically blue
colors. To test various statistical differences between void
galaxies and galaxies from the samples selected using more general
criteria, we compare some of the parameters of void galaxies with
similar data for the sample of 195~galaxies from the Equatorial
Survey (ES) based on a part of the HIPASS blind  H\,I survey. The
compared samples  have similar properties in the common luminosity
range $-18.5 < M_{g} < -13.5$. The faintest void galaxies
differ appreciably from the ES survey  galaxies. However, the ES
survey also contains about  7\% of the so-called ``inchoate\!''
galaxies with high $\M({\rm H\,I})/L_{B}$ ratios,  most of which
are located far from massive neighbors and are probably analogs of
void galaxies.


\end{abstract}

\maketitle

\section{INTRODUCTION}
\label{sec:intro}

Voids in the large-scale distribution of galaxies were discovered
more than 30 years ago
(e.g.,~\cite{joeveer78,kirshner81}).
Observationally they are commonly defined as regions with no
galaxies of normal and high luminosity: \mbox{$M_{B} \geq -20$},
which corresponds to the break in the galaxy luminosity function
(e.g.,~\cite{montero09}). Voids occupy more than
half of the volume of the present-day Universe. Note that the
number of galaxies in the voids does not exceed  20\% of the total
number of catalogued galaxies, and this fraction is indicative of
a significantly lower mass density inside these structures. In
numerical models of the evolution of matter in the hot Universe
with dark matter (DM), voids arise as natural structures and, on
the whole, resemble the observed voids rather closely. However,
the number of galaxies observed in voids is several times smaller
than the predicted number of gravitationally bound DM haloes. The
cause of this phenomenon, which was formulated in different forms
by de~Lapparent~\cite{deLapparent95},
Peebles~\cite{Peebles01}, and Tikhonov and
Klypin~\cite{Tikhonov09}, remains unclear.

It is hoped that further improvement of models of the formation
and evolution of galaxies will reduce the gap between the number
of DM haloes and real galaxies. The above phenomenon may be
caused, among other things, by the higher fraction of low surface
brightness galaxies (LSBD) in the voids, because such objects are
more difficult to detect and identify. More detailed and extensive
studies of the properties of galaxies in the voids are needed to
understand the situation with such objects. Most of the authors
working in this direction
(e.g.,~\cite{Rojas04,Rojas05,patiri06})
studied large distant voids and, because of natural observational
selection, dealt only with the brightest void galaxies  ($M_{r}
\lesssim -16.5$). No substantial differences have been found
between these galaxies and similar galaxies in denser structures,
except that the former have somewhat bluer colors. In their recent
paper Hoyle et al.~\cite{Hoyle12} studied the
photometric properties of almost  90\,000 galaxies in about
1000~voids by analyzing their SDSS images. They similarly found
that dwarfs of all types in voids are systematically bluer
compared to their analogs in denser environments.

One would expect, based on general considerations and simulations,
that the effects of the environment on the evolution should be
most significant for the least massive galaxies
(e.g.,~\cite{kreckel11}). The currently available
results suggest that the following factors should influence the
evolutionary status of low-mass galaxies in the voids.
\begin{list}{}{
\setlength\leftmargin{2mm} \setlength\topsep{2mm}
\setlength\parsep{0mm} \setlength\itemsep{2mm} }

\item (1) An appreciably delayed formation of gravitationally bound DM
halos with masses
typical of dwarf galaxies in low-density large-scale structures
(voids)~\cite{Einasto06} and slow evolution of (a part of) void galaxies.

\item (2) Voids are filled mostly with low-mass galaxies.
Therefore, because of the well-known correlation between
luminosity and surface brightness
(e.g.,~\cite{cross02}), their population contains
a higher fraction of LSB galaxies. Numerical simulations of the
interaction of low- and normal- (high-, HSB) surface brightness
disk galaxies~\cite{mihos97} in the case of their
encounters without mergers show that the final response differs
strongly for the two types of objects. HSB galaxies respond to
such interactions by developing a bar and a mature burst of star
formation, whereas the response in \mbox{LSB} galaxies is much
weaker with the star-formation rate increasing only slightly. The
number of collisions in voids during the lifetime of a galaxy is
several times smaller than in the average-density regions, and
therefore one would expect an appreciable fraction of void
galaxies to have never been affected by ``significant''\
interactions. However, these general considerations should be
validated by numerical simulations taking into account many subtle
details of the  formation and evolution of galaxies.

\item (3) Kreckel et al.~\cite{kreckel11} analyzed the results of such
a simulation of the evolution of galaxies in voids and found  evidence 
suggesting
that these galaxies differ from the galaxies in denser regions, but this 
is only true
for the lowest-mass objects
in their modeled  mass range  (which corresponds to luminosities $M_{r}$ 
ranging from
$-12^{\rm m}$ to $-16^{\rm m}$).
\end{list}

It follows from the above that one must search for the possible evolutionary 
peculiarities in the
least massive void galaxies by studying sufficiently close objects.

The first deep sample of 79 galaxies  (down to $M_{B} \sim-12$) in
the nearby Lynx-Cancer void was presented by Pustilnik and
Teplyakova~\cite{PaperI}. The results of their
study suggest that the effect of low-density environment indeed
exists for such low-mass galaxies, and that it manifests itself in
the slower rate of evolution. According to the conclusions of
Pustilnik et al.~\cite{PaperII}, void galaxies
have systematically lower metallicities  (on the average, by 30\%)
compared to the galaxies in denser environments. Furthermore, a
small but quite significant  (about 10\%) fraction of void
galaxies have peculiar properties typical of evolutionarily young
(unevolved)
objects~\cite{PaperIII,Triplet}.
We should also mention  the so-called Void Galaxy Survey
(VGS,~\cite{kreckel12}) carried out to study the
H\,I structure and optical properties of about 60 SDSS galaxies
residing near the centers of large but relatively distant
(approximately 80~Mpc) voids. Only for some of the closest voids
of this sample, the luminosities of the galaxies studied are more
or less close to the median luminosity of the sample in the
Lynx-Cancer void. It is among such galaxies that Kreckel et
al.~\cite{kreckel12} found three very gas-rich
objects.

In this paper we report the results of our photometric study of
the galaxies in the  Lynx-Cancer void based on their
$u,g,r,i$-band images from the SDSS~DR7 (Sloan Digital Sky Survey
Data Release 7) database~\cite{DR7}. The aim of
our photometric analysis is to determine the basic parameters of
the sample galaxies in order to study the statistical properties
of void galaxies and compare the results with those obtained for
other samples of similar galaxies in denser environments or
samples based on other criteria. The integrated magnitudes and
color indices of the galaxies can be used to estimate their
stellar masses. Furthermore, the colors of the outer regions of
the galaxies, which usually bear no traces of recent or ongoing
star formation, are compared to model evolutionary tracks and used
as age indicators for the oldest (visible) stellar population. One
of the aims of this study is to estimate these ages.

\section{UPDATED GALAXY SAMPLE IN THE LYNX--CANCER VOID}
\label{sec:sample}

A detailed description of the original sample of galaxies in the
Lynx-Cancer void can be found in our earlier
paper~\cite{PaperI}, where we present a list of
79~galaxies residing in this void---a simply-connected domain
containing no galaxies of normal and high luminosity (here
\mbox{$M_{B} < -19.0$}). The void is described by a sphere with a
radius of 8~Mpc together with the adjoining regions. The void
galaxies satisfy the condition of sufficient separation (more than
2~Mpc) from galaxies brighter than \mbox{$M_{B} = -19.0$}. We
discuss the completeness of the sample in the same paper. As of
now, more than 20 new galaxies of the void have been found. Some
of them were discovered as a result of a program of the search for
new \mbox{LSB dwarf} galaxies in this void, and the remaining
ones---as a result of new surveys (mostly the ALFALFA
survey~\cite{haynes11}) and a detailed analysis of
the data for the already known galaxies in this sky area.

We are preparing the updated sample of galaxies of the
Lynx-Cancer void for publication. The new version includes 101
galaxies, 16 of which lie outside the sky area covered by SDSS
fields. We limit our analysis to 85 galaxies listed in Table~1. We
base our study on the updated sample in order to study the
photometric properties of void galaxies with maximum completeness.

\section{METHODS OF PHOTOMETRIC REDUCTION}
\label{sec:method}

The SDSS survey~\cite{York} offers high-quality
calibrated images taken in the $u,g,r,i,z$ bands. Therefore
primary reduction has to include only one extra step---thorough
subtraction of the sky background. We performed this procedure in
the ESO-MIDAS environment using the {\tt aip} package as described
in detail in the paper by Kniazev et
al.~\cite{LSB-SDSS}. The object under study is
masked and the sky background is approximated by a two-dimensional
polynomial. The resulting sky background model is then subtracted
from the image. The sky background in the masked region is
interpolated from the surrounding region. Ring  aperture
photometry is then performed to measure the flux inside the masked
region for the \mbox{$u,g,r,i$-band} images with subtracted
background. The inferred fluxes are then converted into integrated
magnitudes of the galaxies. To convert the instrumental fluxes
into magnitudes, we use a set of photometric coefficients for each
field of the SDSS database.

We determine the following model-independent parameters for all
galaxies: integrated $u,g,r,i$-band magnitudes, integrated
$(u-g),(g-r),(r-i)$ colors, and integrated $B_{\rm tot}$
magnitudes converted from the $g$- and $r$-band magnitudes by the
formulas proposed by Lupton~\cite{Lupton05}. After
creation of the surface brightness profile, we estimate for each
object the radii $R_{\rm 50}$ and $R_{\rm 90}$ containing 50\% and
90\% of the $u,g,r,i$-band fluxes of the galaxy. In addition, we
estimate the effective surface brightness inside the $R_{\rm 50}$
radius, the observed minor-to-major semiaxis ratio $b/a$, the
``optical'' and Holmberg radii~\cite{holmberg} (at
the $\mu(B) = 25.0$ and $26\fm5/\square''$ surface brightness
levels respectively). We multiply all the effective radii  by a
factor of $\sqrt{a/b}$.

Fitting  an exponential or Sersic~\cite{Sersic}
law to the radial surface brightness profiles allowed us to
determine the model parameters $\mu_{\rm 0}$ of the central
$u,g,r,i$-band surface brightness. We then compute the $B$-band
central brightness $\mu_{\rm 0}(B)$  from  $\mu_{\rm 0}(g)$ and
$\mu_{\rm 0}(r)$ by Lupton's
formulas~\cite{Lupton05}. We also estimate the
$u,g,r,i$, and $B$-band ($\mu_{0,c,i}(B)$) central surface
brightness values corrected for the inclination of the galaxy disk
to the line of sight and the foreground extinction in our Galaxy
in accordance with~\cite{SF11}. We compute the
inclination correction by the formula \mbox{$\delta \mu =
-2.5\log(\cos i)$}, where $i$ is the angle between the disk plane
and the line of sight. For objects with inclinations \mbox{$i >
70\degr$}, we compute the corresponding correction by the formula
\mbox{$\delta\mu = 2.5\,\log
({\alpha}/{\alpha}_z)\,\left(1-\exp(-\tan(i)\,{\alpha}/{\alpha}_z)\right)$}~\cite{Matthews99},
where we set the  ${\alpha}/{\alpha}_z$ scale length ratio equal
to~$5$~\cite{Kruit81}. We compute both corrections
assuming negligible  internal extinction in the galaxies
considered. We compute the angle~$i$ by the standard formula
\mbox{$\cos^2\!i = (p^2 - q^2)/(1-q^2)$}, where \mbox{$p=b/a$} and
$q$ are the observed axial ratio and the axial ratio of the real
disk respectively. We set  \mbox{$q = 0.2$} for relatively bright
late-type galaxies and \mbox{$q = 0.4$} for low-luminosity objects
with \mbox{$M_{B}
>-14.8$} in accordance with the results of Roychowdhury et
al.~\cite{Roych13}. We also estimate the
(characteristic) disk scale length $\alpha$ for the
\mbox{$u,g,r,i$-band} images and for the Sersic exponent~$n$.

\subsection{Determination of the Ages\\ of the Old Stellar Population}

We estimate the colors of the outer parts of the galaxies by
performing additional photometry  inside a set of small circular
apertures (with the radii of about~4\arcsec) at the periphery of
the galaxies excluding small higher-luminosity areas. We then sum
up the flux in these apertures and convert it into magnitudes and
colors. For the objects with uncertain positions of the H\,II
regions, we use the  H$\alpha$ images taken with the 6-m telescope
of the Special Astrophysical Observatory. We estimate the time
elapsed since the onset of an epoch or a ``short episode''\ of
star formation, which under certain assumptions can be used as an
age estimate for the oldest visible stars, by comparing the colors
of the peripheral regions of galaxies with model evolutionary
tracks computed using the {\tt PEGASE2} software
package~\cite{pegase}.

When selecting the peripheral areas to determine the colors of the
oldest population, we choose the areas located far from the center
but where the fluxes could still be measured with reasonable
accuracy. As a result, the surface brightness levels of the
considered peripheral areas vary substantially for galaxies with
different surface brightness and/or characteristic size ranging
from ${\rm SB}_B=24\fm5/\square''$ to $26\fm9/\square''$ with a
median value of ${\rm SB}_B({\rm median}) = 25\fm3/\square''$. We
convert the ${\rm SB}_g$ values, actually measured in the radial
surface brightness profiles, into  ${\rm SB}_B$ parameters
adopting the color index \mbox{$B-g=0.4$} typical of void
galaxies. 
\onecolumngrid

\begin{figure}
\setcaptionmargin{5mm} 
\captionstyle{normal}
 \includegraphics[angle=-90,width=0.8\columnwidth]{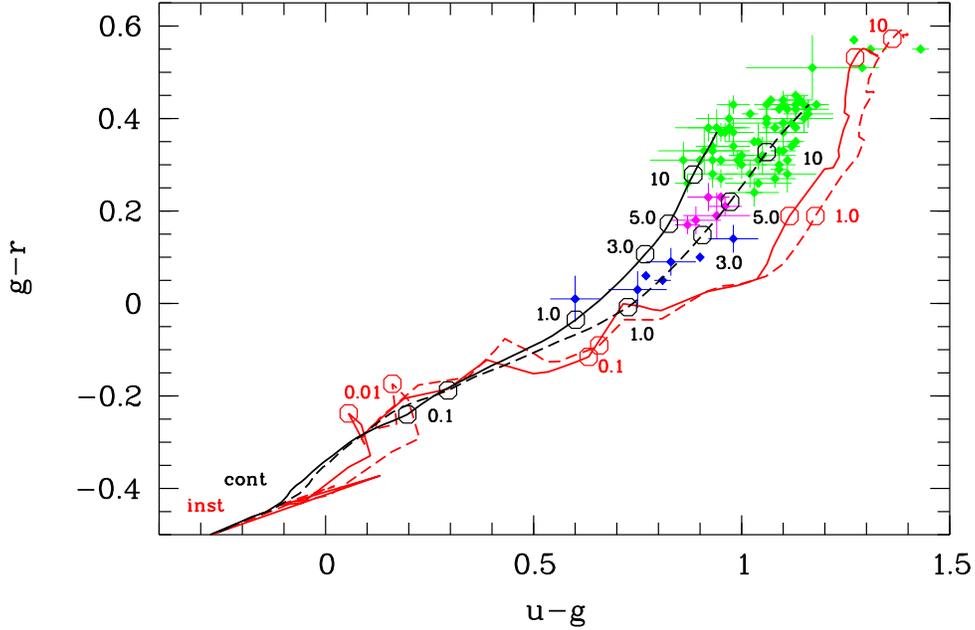}
\vspace{1pt}
  \caption{\label{fig:ugr_periph}
The  $(u-g)$--$(g-r)$ color--color diagram of the outer parts of
85 galaxies of our sample of the  Lynx-Cancer void compared to
{\tt PEGASE2} evolutionary tracks for two extreme cases of the
star-formation law. Instantaneous ($\rm inst$) SF for two initial
mass functions (IMF): Salpeter (the red solid track) and Kroupa (the
red dashed track). Continuous ($\rm const$) star formation for
the same two IMFs: Salpeter (the black solid track) and Kroupa (the
black dashed track). The green, pink, and blue circles
show the colors of the outer parts of the galaxies corresponding
to the ages \mbox{$T\sim10$--$13$~Gyr},
\mbox{$T\sim4$--$6.5$~Gyr}, and \mbox{$T\sim1$--$3.5$~Gyr}
respectively. The numbers near each track indicate the time in Gyr
elapsed since the onset of star formation. }
\end{figure}

\begin{figure}
\setcaptionmargin{5mm} 
 \includegraphics[angle=-90,width=0.8\columnwidth]{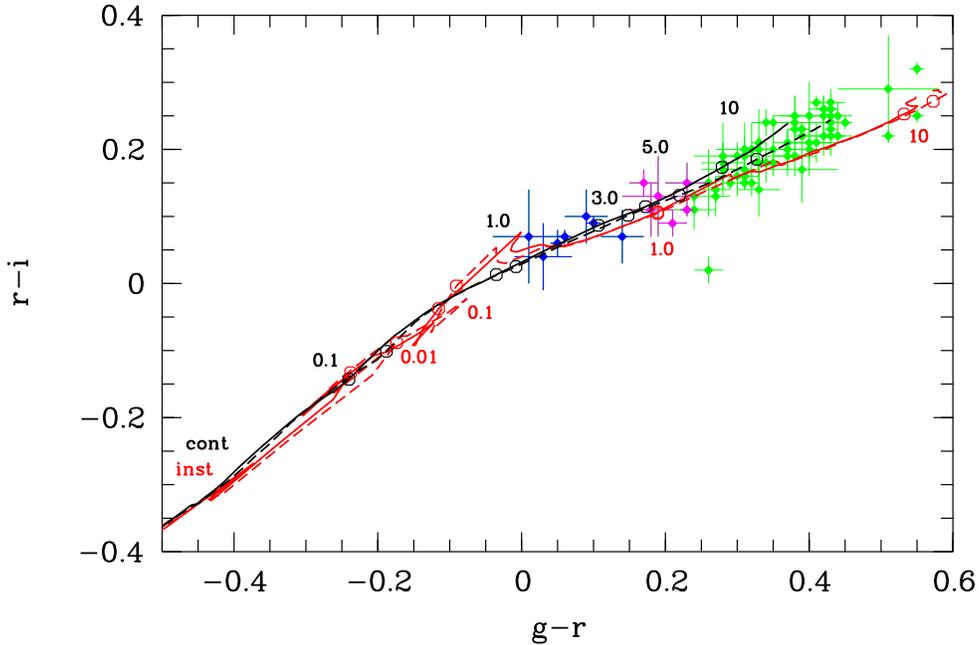}
\caption{\label{fig:gri_periph} The
$(g-r)$--$(r-i)$ color-color diagram of the outer parts of  85
galaxies in the Lynx-Cancer void compared to  {\tt PEGASE2}
evolutionary tracks for two extreme cases of the star formation
law (same designations as in~Fig.~\ref{fig:ugr_periph}).}
\end{figure}

\newpage
\twocolumngrid

The {\tt PEGASE2} software package is designed for computing
models of the photometric evolution of galaxies. The main
parameters of the models are: the initial mass function
$\varphi(\M)=\M^{-(1+x)}$, the star-formation law ${\rm SFR}(t)$,
and stellar metallicity $Z$. We use for our analysis the initial
mass functions (IMF) of Salpeter~\cite{Salpeter}
and Kroupa et al.~\cite{kroupa}, which best
describe the observed distributions of stellar masses in the
nearby galaxies. We use two extreme scenarios for the
star-formation law: an instantaneous episode---a burst---of star
formation ($\rm inst$) and star formation at a constant rate ($\rm
const$). The tracks of all other non-exotic scenarios lie between
these two extreme cases. For any fixed time since the onset of
star formation, the $\rm const$ scenario track produces the bluest
colors, because the fraction of massive blue stars for this track
decreases at a slower rate than the corresponding fraction for the
$\rm inst$ scenario track. In the domains of the $ugr$ diagram
where the $\rm const$ and $\rm inst$ tracks run sufficiently close
to each other (see Fig.~\ref{fig:ugr_periph}), the
time elapsed since the onset of star  formation and,
correspondingly, the ages of the oldest stars are about five times
greater for the $\rm const$ track  than the  corresponding
quantities for the $\rm inst$ tracks. As a result, in the cases of
uncertain estimates, the $\rm const$ tracks yield robust upper
limits for the estimated ages of the oldest stars, which therefore
can be used as the upper estimates for the ages of the visible
extended old population. We show all our inferred colors of the
outer regions of the program galaxies in the   $ugr$ and $gri$
two-color diagrams, where we compare these colors with the model
tracks described above
(Figs.~\ref{fig:ugr_periph}~and~\ref{fig:gri_periph}).

When comparing the observed colors of the galaxies with those of
model tracks, we choose the stellar metallicity for the computed
tracks from a discrete set of values offered by {\tt PEGASE2}
adopting the one that is the closest to the measured gas
metallicity for a particular galaxy. In most cases we use one of
the following three metallicity values: $Z=0.004$, $Z=0.002$, or
$Z=0.001$.

The  filter transmission curves used to compute the {\tt PEGASE2}
evolutionary tracks slightly differ from the transmission curves
of the $u,g,r,i$ filters used in actual SDSS observations. We use
the formulas from Tucker et al.~\cite{Tucker06} to
eliminate the resulting small biases.

\subsection{Gas Mass Fraction}

The total hydrogen mass $\M({\rm H\,I})$ for the galaxies of our
sample with measured integrated \mbox{21-cm} HI~line fluxes
$F({\rm H\,I})$ can be estimated using the formula for the
optically thin layer: \mbox{$\M({\rm H\,I})=2.36\,\times$}
\mbox{$F({\rm
H\,I})\,D^{2}\,10^{5}$~\cite{Roberts69}}, where
$D$ and $F({\rm H\,I})$ are the distance to the galaxy in Mpc and
flux in Jy~km~s$^{-1}$ respectively. One must also take into
account the helium contribution ($0.33$ of $\M({\rm H\,I})$) to
estimate the total gas mass $\M_{\rm gas}$. The contribution of
the mass of molecular gas in dwarf galaxies can be neglected in
the first approximation. For about half of the galaxies, we adopt
the $F({\rm H\,I})$  data  from the literature and determine the
corresponding fluxes for the remaining galaxies from observations
performed with the NRT radio telescope.

The photometry obtained as a result of this study can be used to
estimate the total stellar masses. Given the gas mass and that of
the stars, an evolutionary parameter---the gas mass fraction---can
be determined. The total stellar mass $\M_{*}$ is commonly
estimated from the integrated photometry of a galaxy in one of the
broadband filters  $\lambda$ using the corresponding parameter
$\Upsilon_{\lambda} = \M_{*}/L_{\lambda}$---the mass-to-luminosity
ratio for the stellar population, which depends on the color of
the stellar continuum. We use for our estimates of $\M_{*}$ the
mass-to-luminosity ratio  $\Upsilon$ and its dependence on the
color indices as derived by Zibetti et
al.~\cite{Zibetti09}. More precisely, we use the
ratio $\Upsilon_{g}(g-i)$ and  $g$-band luminosity, because this
combination yields the most robust results in the optical part of
the spectrum. In our opinion, the mass-to-luminosity ratio
$\Upsilon$ from~\cite{Zibetti09} takes most adequately into
account the complex history of star
formation in low-mass galaxies including the recent episodes.

We use this stellar mass estimate $\M_{*}$ to compute the baryonic
mass \mbox{$\M_{\rm bary}=\M_{*}+\M_{\rm gas}$} of the galaxy and
then the gas mass fraction $f_{\rm gas}$, i.e., the ratio of the
total gas mass to the total baryonic mass. To compare our results
with those obtained for other samples, we also compute the
model-independent parameter $\M({\rm H\,I})/L_{B}$ which is the
ratio of the hydrogen mass to $B$-band luminosity in solar units;
this parameter is often used in less detailed studies. Here
$L_{B}$ is determined from the $B_{\rm tot}$ magnitude computed
using Lupton's formulas~\cite{Lupton05} from our
independent $g$ and $r$-band photometry. This parameter provides a
coarse estimate of the gas mass fraction for galaxies with no
color information available.

\section{RESULTS}
\label{sec:results}

Table~1 lists the basic parameters for our sample of   \mbox{85}
galaxies of the Lynx-Cancer void. The columns in the table give:
(1)~the name of the galaxy; (2)~the name of the object in the
short IAU format; (3)~type of the galaxy  either adopted
from~\cite{PaperI} (for the objects from the
original
sample), or NED,\!\footnote{NASA/IPAC Extragalactic Database,\\
{\tt http://ned.ipac.caltech.edu}} or, if such information is
unavailable in NED, according to our own estimate;
(4)~heliocentric radial velocity $V_{\rm hel}$ in km~s$^{-1}$;
(5)~distance to the object in Mpc (estimated as
in~\cite{PaperI}, i.e., corrected for the large
negative peculiar velocity in the region considered); (6)~Galactic
$B$-band extinction $A_{B}$ adopted
from~\cite{SF11}; (7,~8)~the total apparent
(uncorrected for Galactic extinction) magnitudes $B_{\rm tot}$ and
the absolute (extinction-corrected) magnitudes $M_{B,0}$ derived
from our independent photometry by Lupton's
formulas~\cite{Lupton05} and used in the
subsequent statistical analysis; (9)~total apparent  $B_{\rm tot}$
magnitudes adopted from NED (for comparison with our magnitude
estimates); (10)~hydrogen mass $\M({\rm H\,I})$ in the units of
$10^7\,M_{\odot}$ (according to the published data and the results
of observations made with the NRT radio telescope); (11)~hydrogen
mass to $B$-band luminosity ratio $\M({\rm H\,I})/L_{B}$ in solar
units; (12)~stellar mass $\M_{*}$ in the units of
$10^7~M_{\odot}$; (13)~gas mass fraction $f_{\rm gas}$;
(14)~oxygen abundance $12+\log({\rm O/H})$; (15)~the observed
axial ratio $b/a$. 

Model-independent parameters derived from photometry and SDSS image
analysis
are given in Tab.~2 and 3. In columns of Tab.~2 the
following parameters are presented:
(2),(3) the radii
containing 50\% and 90\% of the  $g$-band flux of the galaxy;
(4) effective $g$-band surface brightness determined as the average
surface brightness inside the $R_{\rm 50}(g)$ radius; (5),(6) $g$-band
``optical''\ radius (at the $25\fm0/\square\arcsec$ isophote
level) in arcsec ($R^{25}(g)$) and kpc~($a_{\rm 25}$); (7)  $g$-band
magnitude inside the ``optical''\ radius; (8),(9) radius at the
$26\fm5/\square\arcsec$ isophote level
\begin{figure}
\setcaptionmargin{5mm} \onelinecaptionsfalse
\captionstyle{normal} \captionstyle{normal}
 \includegraphics[angle=-90,width=\columnwidth,bb=0 25 539 765,clip]{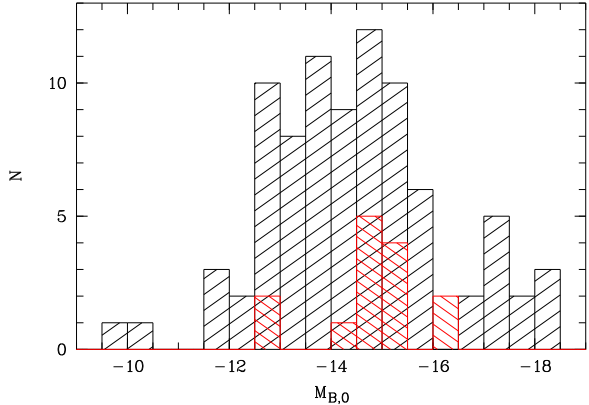}
\caption{\label{fig:hist_MB} Distribution of the
$B$-band absolute magnitudes for 85 galaxies of the Lynx-Cancer
void. The red histogram shows the distribution for the
subsample of the 14 so-called  ``inchoate'' galaxies
from~\cite{Garcia09}.}
\end{figure}
(the Holmberg
radius)~\cite{holmberg} and the $g$-band magnitude
inside this radius; (10),(11) optical size computed as \mbox{$b =
R^{25}(g)\,\sqrt{b/a}$} and \mbox{$a = R^{25}(g)\,\sqrt{a/b}$} in
arcsec and kpc respectively.
In columns of Tab.~3 the following parameters are oresented:
(2) integrated $g$-band magnitude;
(3),(4),(5)
Galactic extinction-corrected integrated colors \mbox{$(u-g)$},
$(g-r)$, $(r-i)$; (6),(7),(8) the colors of outer parts
\mbox{$(u-g)$}, $(g-r)$, $(r-i)$.

Fitting the photometric profiles to an exponential
disk or, in a more general case, to Serscic's
law~\cite{Sersic} yields the model parameters that
are also listed in Table~4: (2),(3) the central $g$ and $r$-band
surface
brightness; (4),(5) the central $B$-band surface brightness values (computed
by Lupton's formulas~\cite{Lupton05}) uncorrected
and corrected for Galactic extinction and inclination to the line
of sight; (6) the scale length in arcsec; (7) the Sersic index. \\

\begin{figure}
\setcaptionmargin{5mm} \onelinecaptionsfalse
\captionstyle{normal}
 \includegraphics[angle=-90,width=0.98\columnwidth,bb=0 25 539 765,clip]{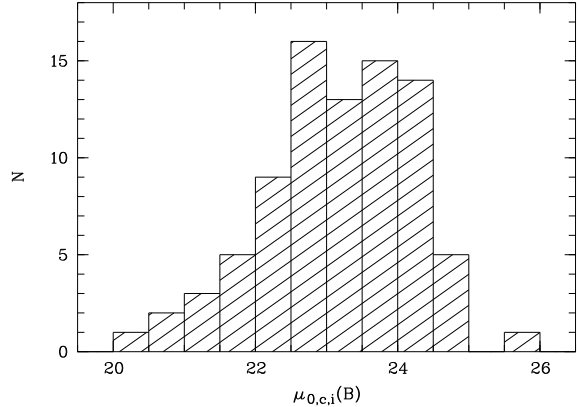}
 \vspace{3pt}
\caption{\label{fig:hist_muBOi} Distribution of
the $B$-band central surface brightness corrected for Galactic
extinction and inclination to the line of sight for 85 void
galaxies. }
\end{figure}

\section{ANALYSIS AND DISCUSSION}
\label{sec:dis}

\subsection{Parameter Distributions}

We construct the observed distributions of the most important
quantities for the sample studied to understand the full range of
their variation. Figure~\ref{fig:hist_MB} (the
sparse hatching)  shows the distribution of absolute magnitudes
$M_{B,0}$ for the void objects  computed from our photometry and
corrected for Galactic extinction. For comparison, we show the
distribution of $M_{B,0}$  for the subsample of the so-called
``inchoate'' galaxies from the ES
survey~\cite{Garcia09} (red hatching in the
same figure). 
\begin{figure*}
\begin{minipage}{0.49\linewidth}
\setcaptionmargin{5mm} \onelinecaptionsfalse
\captionstyle{normal}
 \includegraphics[angle=-90,width=\columnwidth]{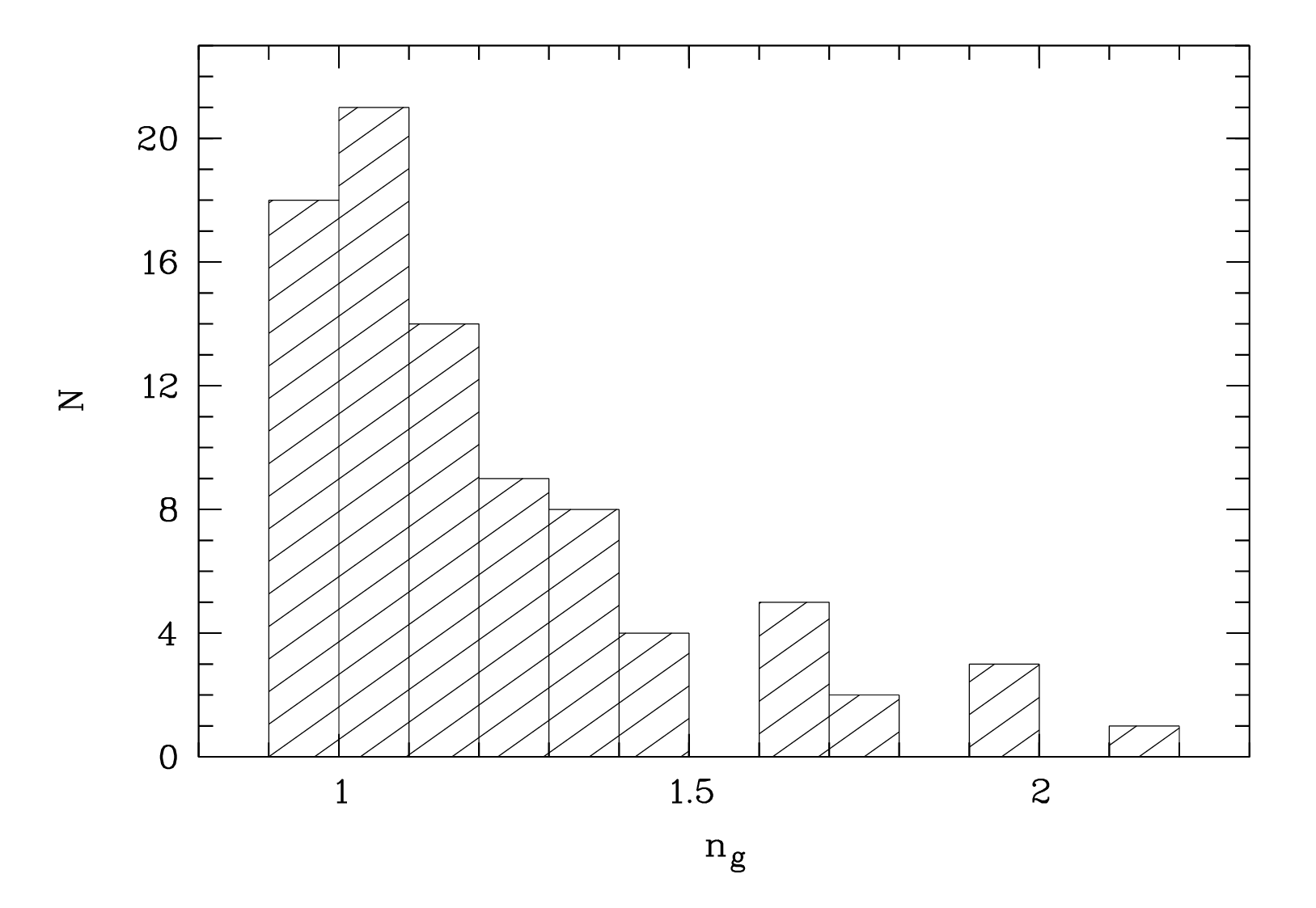}
\caption{\label{fig:hist_Ng} Distribution of the
Sersic index~\cite{Sersic} of the radial
\mbox{$g$-band} surface brightness profile for 85 void galaxies. }
\end{minipage}
\begin{minipage}{0.49\linewidth}
\setcaptionmargin{5mm} \onelinecaptionsfalse
\captionstyle{normal} \vspace{0mm}
  \includegraphics[angle=-90,width=0.98\columnwidth]{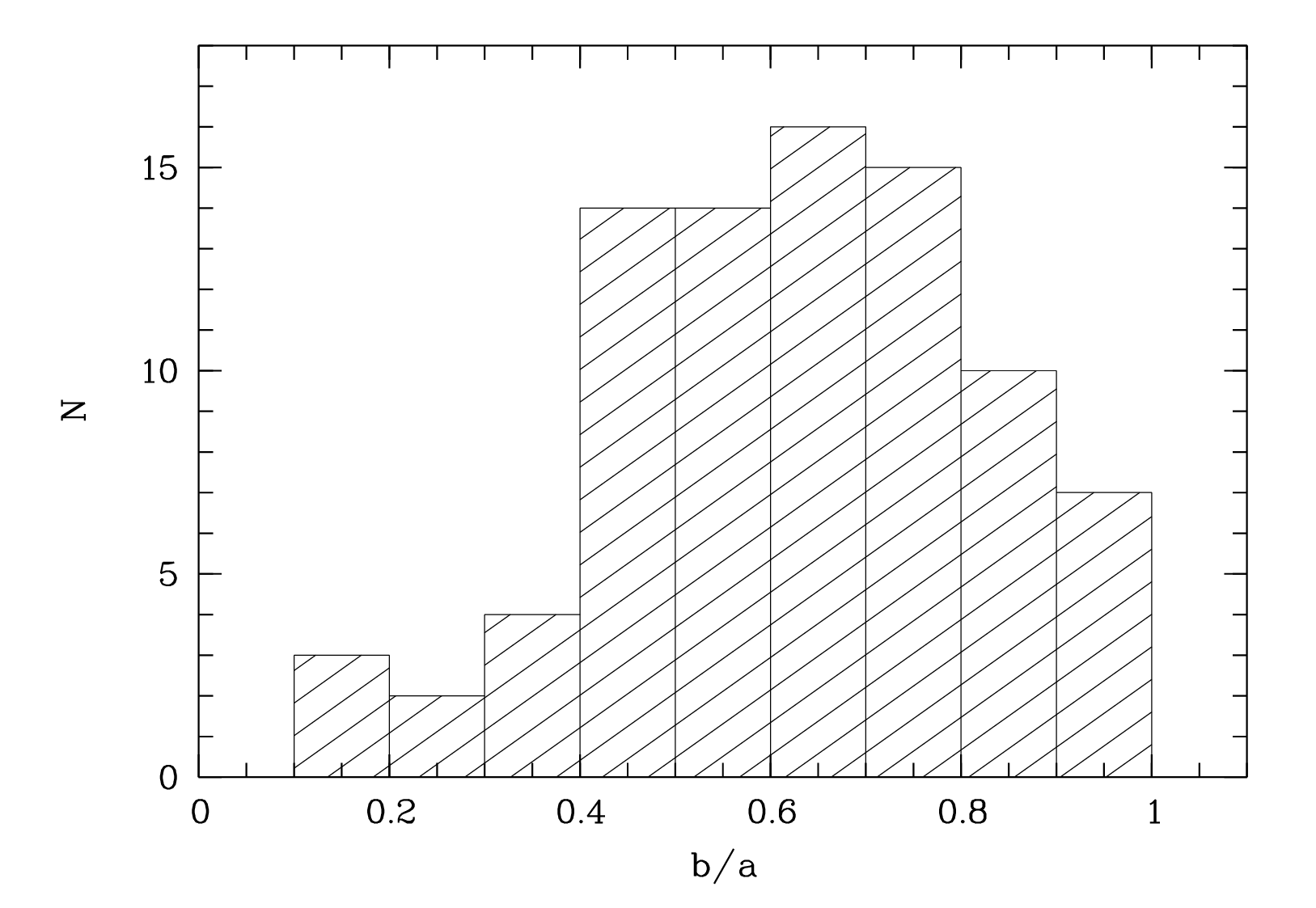}
 \vspace{6mm}
\caption{\label{fig:hist_b_a} Distribution of the
observed axial ratio $b/a$ for 85~void galaxies.}
 \end{minipage}
\end{figure*}

Garcia-Appadoo et
al.~\cite{Garcia09} selected these  14 galaxies
from the full list of 195 Equatorial-Survey objects on the basis
of their peculiar properties - blue colors and high $\M({\rm
H\,I})/L_{g}$ ratios. The above authors overestimate the 21-cm
line fluxes for some of these objects which leads to overestimated
$\M({\rm H\,I})/L_{g}$ ratios. We have corrected these values by
taking into account the new published data, and the recomputed
ratios are no longer so outstanding. In particular, the $\M({\rm
H\,I})/L_{B}$ ratio for the NE component of the H\,I\,1225+01
(HIPEQ\,1227+01) pair is now equal to $10.1$ instead of the
original estimate of $22$. In addition, to better understand the
nature of the galaxies in this subsample, we searched their
neighborhoods for the presence of massive galaxies. Of the 14
``inchoate'' galaxies nine are separated by at least 1~Mpc from
the closest massive galaxy, and for the six most gas-rich objects
the distances to the closest massive galaxies lie in the range of
\mbox{2--5 Mpc}. Thus, the galaxies of this subsample,
like our void galaxies, show evident signs of isolation.
\begin{figure*}
\begin{minipage}{0.49\linewidth}
\setcaptionmargin{5mm} \onelinecaptionsfalse
\captionstyle{normal}
 \includegraphics[angle=-90,width=\columnwidth]{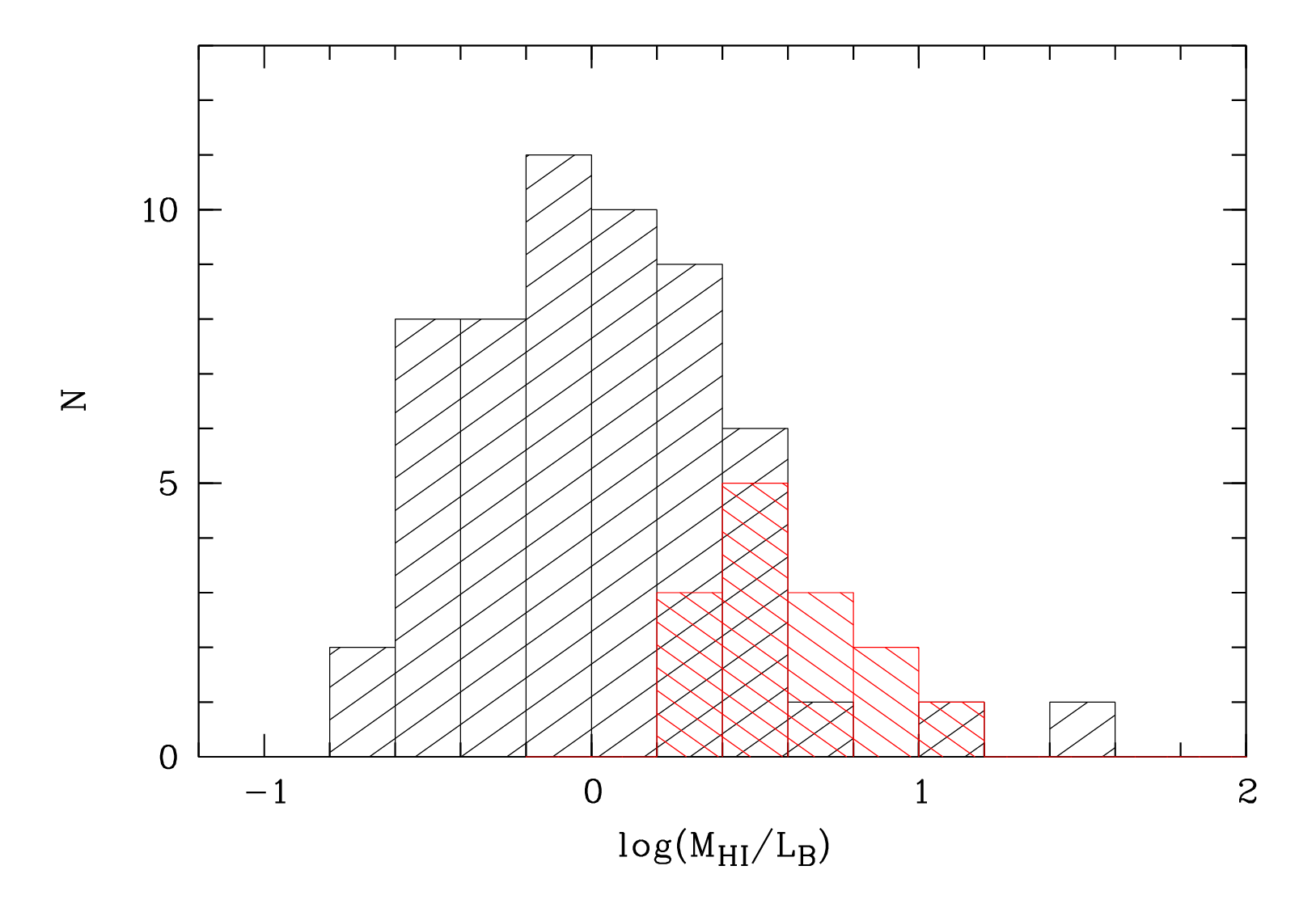}
\caption{\label{fig:hist_MHI_LB} Distribution of
the hydrogen mass to $B$-band luminosity ratio $\M({\rm
H\,I})/L_{B}$ (in solar units) for the void galaxies. The
red histogram shows the distribution for the
subsample of  14 ``inchoate''\ galaxies from the total sample of
the blind  H\,I survey (ES) from~\cite{Garcia09}.
}
\end{minipage}
\begin{minipage}{0.49\linewidth}
\setcaptionmargin{5mm} \onelinecaptionsfalse
\captionstyle{normal} \vspace{-4mm} \hspace{-2mm}
 \includegraphics[angle=-90,width=\columnwidth]{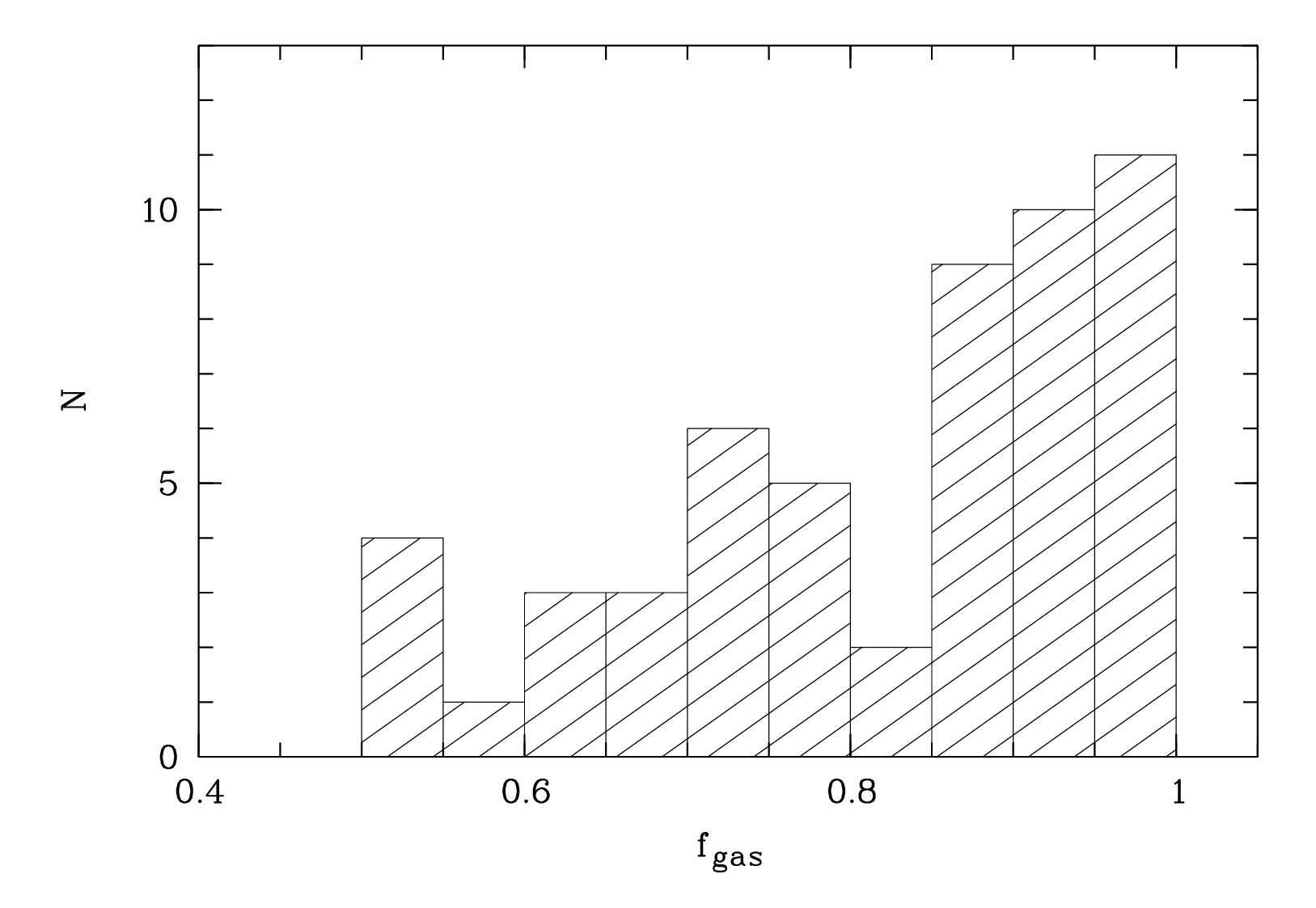}
\vspace{0mm}
 \caption{\label{fig:hist_mu_gas}
Preliminary distribution of the gas mass fraction for about half
of the sample objects (54 void galaxies) based on published H\,I
data and observations made with the NRT radio telescope (see text
for details).
  }
\end{minipage}
\end{figure*}

Figure~\ref{fig:hist_muBOi} shows the distribution
of the $B$-band central surface brightness $\mu_{0,c,i}(B)$
(corrected for Galactic extinction and inclination to the line of
sight) for the studied subsample of the Lynx-Cancer\linebreak
void galaxies. The fraction of galaxies with\linebreak
\mbox{$\mu_{0,c,i}(B)>23\fm0/ \square\arcsec$} (LSB) in this
subsample is equal to about 50\%.
Figure~\ref{fig:hist_Ng} shows the distribution of
the Sersic index $n_{g}$ determined by modeling the radial
$g$-band brightness profile. The index $n_{g}$ is sufficiently
close to $1$ for almost two thirds of the objects, i.e., they have
close-to-exponential brightness profiles. The brightness profiles
of the remaining galaxies with the indices $n_{g}$ in the range of
1.2 to 2.1 flatten out appreciably towards the center.
The distribution of the observed axial ratios $b/a$ in
Fig.~\ref{fig:hist_b_a} shows that about 50\% of
the sample objects are tilted significantly to the line of sight
($b/a<0.65$ and, correspondingly,  $i>50\degr$). As we showed in
our previous paper~\cite{PaperI}, because of the
selectivity of the SDSS spectroscopic survey with respect to
surface brightness many  LSB galaxies in the void that are
brighter than the formal SDSS cutoff threshold \mbox{$r_{\rm
petro}< 17.77$}~\cite{Petro} could lack velocity
measurements. Because of the tilt to the line of sight, their
apparent brightness is overestimated, making it more likely for
them to be included in the spectroscopic part of the SDSS survey.

Figure~\ref{fig:hist_MHI_LB} (the sparse hatching) shows the 
distribution
of the hydrogen mass to luminosity ratio $\M({\rm H\,I})/L_{B}$. 
The median value for  57 galaxies of
our sample is $\M({\rm H\,I})/L_{B}\sim1.0$. Here, like in Fig.~\ref{fig:hist_MB},
we show the corresponding distribution for the ``inchoate''\ galaxies 
from~\cite{Garcia09}\
(the red hatching). The median  $\M({\rm H\,I})/L_{B}$ ratio for 
this sample is equal to about $3.4$.
Garcia-Appadoo et al.~\cite{Garcia09}\ associate the term ``inchoate''
with the irregular shapes of these galaxies: the above authors assume 
that these systems are
in the process of formation or at the beginning of their evolution. We 
show below that some of the
galaxies in our sample also have peculiar properties and resemble 
``inchoate'' objects from
the ES sample (see Table~1). Furthermore, they are unusually blue 
\mbox{($(B-V)<0.3$)} and the
median value of their parameter $\M({\rm H\,I})/L_{B} \sim2.7$ is comparable
with the corresponding parameter for the subsample of
Garcia-Appadoo et al.~\cite{Garcia09}.

Figure~\ref{fig:hist_mu_gas} shows the
distribution of the gas mass fraction $f_{\rm gas}$. The gas mass
fraction for  59\% of the void galaxies with available  H\,I data
lies in the range of \mbox{80--99\%}. Such a high concentration of
gas-rich objects in our sample indicates, as we already pointed
out in our earlier paper~\cite{PaperII}, that the
void population evolves more slowly than the galaxies in denser
environments.

\begin{figure}
\setcaptionmargin{5mm} \onelinecaptionsfalse
\captionstyle{normal}
 \includegraphics[angle=-90,width=0.98\columnwidth]{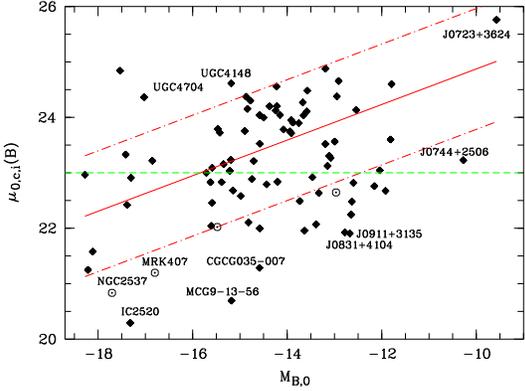}
\caption{\label{fig:MBo_muBOexti} Relation between
the absolute magnitude $M_{B,0}$ and central surface brightness
$\mu_{0,c,i}(B)$ for 85 galaxies of the Lynx-Cancer void. The
horizontal green dashed line corresponds to the surface brightness level
of $23^{\rm m}/\square''$, above which lies the domain of LSB
galaxies. Linear regression (the red solid line) is performed based
on the data points for 81 galaxies of the subsample considered.
The red dash-dotted lines show the scatter of the subsample data
points. Four  BCGs (the open circles) were not used in the
regression (see text). }
\end{figure}
\subsection{Relations Between the Parameters\\ of Void Galaxies}

In Fig.~\ref{fig:MBo_muBOexti} we compare the
absolute magnitude $M_{B,0}$ and the central surface brightness
$\mu_{0,c,i}(B)$ for the sample of 85 galaxies of the Lynx-Cancer
void. Both parameters are corrected for Galactic extinction. The
central surface brightness values are corrected for the galaxies'
inclination to the line of sight. We estimate the central surface
brightness values for galaxies with a substantial contribution
from  the central star-forming region and for galaxies with a
``bulge'' by extrapolating to the center of the underlying
external disk profile. About half of the galaxies are classified
as  LSB, i.e., their corrected surface brightness is
$\mu_{0,c,i}(B) > 23^{\rm m}/ \square\arcsec$ (lies above the
green dashed line). Linear regression (the red solid line) is based on all
the galaxies of the subsample except for four blue compact
galaxies (BCG) with ongoing star formation. The standard deviation
of individual galaxies from the general trend is rather large:
$\sigma_{\mu_{0}} = 0\fm97/ \square\arcsec$ (the red dash-dotted
lines). The open circles show the positions of the four BCGs.
These rare low-mass galaxies have a higher central surface
brightness and differ appreciably from the more typical late-type
dwarf galaxies. 

\begin{figure}
\setcaptionmargin{5mm} \onelinecaptionsfalse
\captionstyle{normal}
 \includegraphics[angle=-90,width=1\columnwidth]{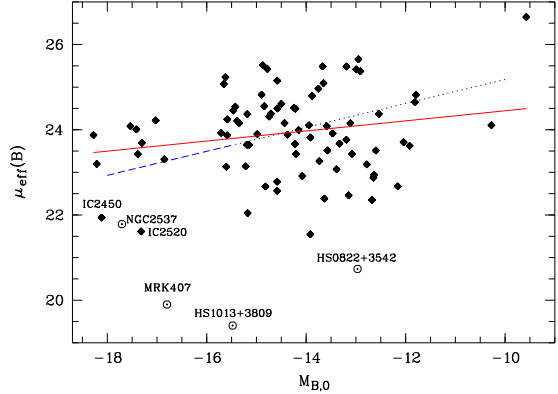}
\vspace{-5pt}
\caption{\label{fig:MBo_muBeff} Relation between
the absolute magnitude $M_{B,0}$ and effective surface brightness
$\mu_{\rm eff}(B)$ for 85 galaxies of the Lynx-Cancer void.
Linear regression (the red solid line) is based on all galaxies except
for the four BCGs (the open circles). A similar line from Cross
and Driver~\cite{cross02} (the blue dashed line) based
on a sample of about 45\,000 galaxies and extrapolated into the
domain of low luminosities down to $M_{B,0}< -10$ (the blue dotted line)
runs close to our relation in the common luminosity range
$M_{B,0} <-15.5$. }
\end{figure}

Figure~\ref{fig:MBo_muBeff} shows the relationship
between the absolute magnitude $M_{B,0}$ and the effective surface
brightness $\mu_{\rm eff}(B)$. Both parameters are corrected for
Galactic extinction. The red solid line shows the linear regression
based on 81 void galaxies except for the four BCGs (its slope is
equal to $k=0.12\pm0.06$). The trend cannot be inferred reliably
because of the large scatter of data points in the central part of
the range. A similar  line (shown by the blue dashes) based on the data for
an extensive sample of galaxies (about 45\,000~objects)  from
Cross and Driver~\cite{cross02} and extrapolated
into the domain of low luminosities (the blue dotted line) runs close
to the center of the distribution for void galaxies but has a
significantly higher slope ($k=0.28$). This difference must be due
to the higher dwarf fraction in the sample of void galaxies and
greater diversity of their properties. Furthermore, the scatter
around the regression line for the sample of Cross and
Driver~\cite{cross02} is smaller than for void
galaxies: $\sigma_{\mu} \sim0\fm5/\square\arcsec$ and
$1\fm0/\square\arcsec$ respectively.

\begin{figure}
\setcaptionmargin{5mm} \onelinecaptionsfalse
\captionstyle{normal}
 \includegraphics[angle=-90,width=\columnwidth]{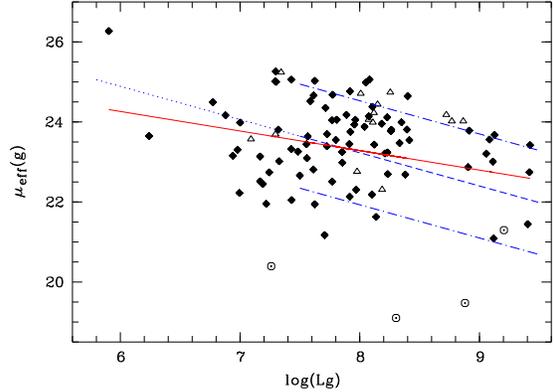}
\caption{\label{fig:logLg_mu_eff} Relation between
the effective surface brightness $\mu_{\rm eff}(g)$ and  $\log
L_{g}$. The red solid line shows the linear regression based on the
galaxies of our sample (the filled diamonds) except for the  BCG
galaxies (the open circles). The blue dashed line shows the linear
regression from Garcia-Appadoo et
al.~\cite{Garcia09}, the blue dotted line shows its
extension to the domain of low luminosities, the blue dash-dotted lines
show the scatter of the data points of the sample of
Garcia-Appadoo et al.~\cite{Garcia09}, and the
open triangles show the group of 14 so-called  ``inchoate''\
galaxies pointed out by the above authors. }
\end{figure}

Figure~\ref{fig:logLg_mu_eff} shows the relation
between the effective surface brightness $\mu_{\rm eff}(g)$ and
luminosity $\log L_{g}$ similar to the relation shown in the
previous figure. Like in the previous figure, linear regression
(the red solid line) is based on all galaxies except for the four
BCGs. To compare our relations with similar relations based on the
sample of 195 galaxies of the equatorial survey
(ES)~\cite{Garcia09} selected by the 21-cm line
emission recorded in the blind HIPASS \mbox{H\,I} survey, we show
the linear regression for this sample  in
Fig.~\ref{fig:logLg_mu_eff} by the blue dashed line. We
extend this line to the domain of low luminosities, where we show
it as a blue dotted line for display purposes. The open triangles
indicate the 14 objects from the subsample of the so-called
``inchoate''\ galaxies of Garcia-Appadoo et
al.~\cite{Garcia09}, which have blue colors and
higher-than-usual hydrogen mass-to-luminosity ratios
(\mbox{$(B-V)\le0.3$} and \mbox{$\M({\rm H\,I})/L_{B}>1.8$}). 
The slope of our regression for void galaxies \mbox{($k
=-0.35\pm0.16$)} is about 2.5~times lower than the corresponding
slope for the  ES-sample \mbox{($k =-0.83$)}. Note that the sample
of Garcia-Appadoo et al.~\cite{Garcia09} contains
no objects with \mbox{$\log L_{g} <  7.5$}. The slope of the
regression based on the ES sample is rather close to the slope for
the sample from~\cite{cross02}---namely, $k=-0.70$
if considering the correlation with the luminosity instead of the
absolute magnitude. The scatter of the parameter $\mu_{\rm
eff}(g)$ for void galaxies is, like the corresponding scatter for
the sample of the ES survey, quite significant, reaching 3 or more
magnitudes at a given luminosity.

\begin{figure}
\setcaptionmargin{5mm} \onelinecaptionsfalse
\captionstyle{normal}
 \includegraphics[angle=-90,width=\columnwidth]{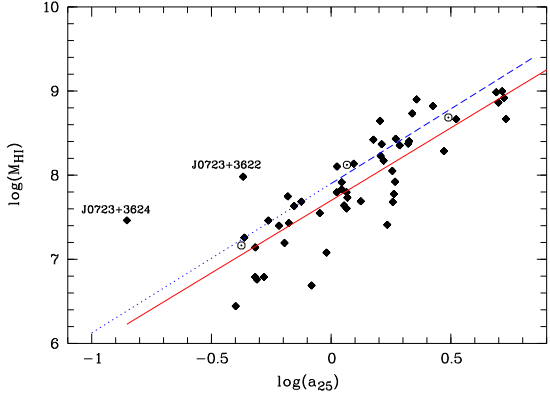}
\caption{\label{fig:Ropt_logMHI} Relation between
the optical radius $\log a_{\rm 25}(g)$ (kpc) and hydrogen mass in
the galaxy $\log\M({\rm H\,I})$ (in solar units). The red solid line
shows the linear regression between these quantities  (its slope
is equal to $1.96$) computed without the BCG galaxies shown by the
open circles. For comparison, we show the relation
from~\cite{Garcia09} (the blue dashed line) and its
extension to the  $a_{\rm 25}(g)<1$ kpc domain (the blue dotted line).
}
\end{figure}

The galaxies at the extremes of the range make the decrease of the
surface brightness with decreasing luminosity in
Figs.~\ref{fig:MBo_muBeff}~and~\ref{fig:logLg_mu_eff}
more or less apparent. However, the rather large scatter of
$\mu_{\rm eff}(g)$ for the void galaxies at the center of the
luminosity range indicates that voids are characterized not only
by the general trend for the increase of the fraction of LSB
galaxies but also by the presence of galaxies with a sufficiently
high surface brightness which must be due to the enhanced star
formation in the last several billion years. The large difference
between the slopes of the regression for the galaxies of the
Lynx-Cancer void and the ES sample may be due to the fact that
the latter contains a substantial fraction of massive galaxies
with enhanced light concentration at the center---a bulge or
traces of a recent burst of star formation. This results in a
shift of the average effective surface brightness at high
luminosities in the ES sample and a larger regression slope.

\begin{figure}
\setcaptionmargin{5mm} \onelinecaptionsfalse
\captionstyle{normal}
 \includegraphics[angle=-90,width=\columnwidth,bb=0 25 539 765,clip]{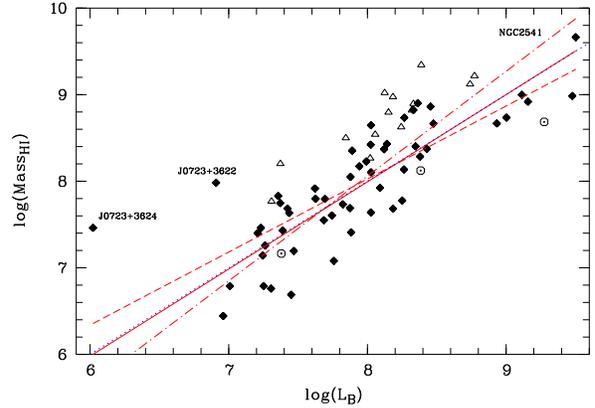}
\caption{\label{fig:logLB_logMHI} Relation between
the hydrogen mass $\log\M({\rm H\,I})$ and luminosity $\log L_{B}$
in solar units. The red dashed line shows the linear regression
throughout the entire luminosity range for the galaxies of our
sample (the filled diamonds), the red solid line shows the linear
regression computed without  J0723+3622 and J0723+3624, and the red
dash-dotted line shows the linear regression over the \mbox{$\log
L_{B}= 7.0$--$8.6$} luminosity interval. All regressions were
computed without BCG galaxies, which are shown by open circles.
The open triangles show the group of the so-called ``inchoate''\
galaxies from~\cite{Garcia09} which stand out
because of their higher gas content. The blue dotted line corresponds
to $\M({\rm H\,I})/L_{B}=1$. }
\end{figure}

Figure~\ref{fig:Ropt_logMHI} shows the relation
between the optical radius $a_{\rm 25}(g)$ (kpc) and hydrogen mass
of the galaxy $\M({\rm H\,I})$ (in solar units). In addition to
the linear regression based on 85 galaxies of our sample (the red
solid line, $k=1.96\pm0.16$), we also show a similar relation
from~\cite{Garcia09} (the blue dashed line). They can
be seen to agree well with each other over the common range of
optical radii. Our data allow this relation to be extended to the
domain of objects with \mbox{$a_{\rm 25}(g) < 1$~kpc}. The
regression slope, which is close to $2$, reflects the simple
physical fact that in the studied galaxies hydrogen is distributed
throughout a rather flat disk and  the characteristic radius of
the H\,I disk closely correlates with that of the optical disk
over a broad range of effective sizes. At the same time, the two
objects (J0723+3624 and J0723+3622) with the highest hydrogen mass
to $B$-band luminosity ratios $\M({\rm H\,I})/L_{B}>10$ deviate
strongly from the main trend.

Figure~\ref{fig:logLB_logMHI} shows the relation
between the hydrogen mass $\M({\rm H\,I})$ and $B$-band luminosity
$L_{B}$ (in solar units) for 54 void galaxies with known
H\,I~masses. The red dashed line shows the linear regression based on
the galaxies of our sample (diamonds, \mbox{$k1 =0.84\pm0.08$}).
The open triangles show the ``inchoate''\ galaxies
from~\cite{Garcia09} for comparison. The blue dotted
line, where $\M({\rm H\,I})/L_{B}=1$, separates the galaxies with
a high gas fraction from the more typical galaxies with normal gas
content. We use the same data to determine the dependence of the
parameter $\M({\rm H\,I})/L_{B}$ on $L_{B}$, whose regression
slope is equal to \mbox{$-0.16\pm0.08$}. This slope is, within the
rather large quoted errors, close to the slope of a similar
dependence found by Pustilnik et
al.~\cite{Pustilnik02}  for BCGs in the Local
Supercluster and in the voids.

A closer look at the figure discussed here suggests that the
increase of  $\M({\rm H\,I})/L_{B}$ with decreasing $L_{B}$ is due
mostly to the galaxies at the extremes of the range, including the
two most gas-rich galaxies of the  J0723+36 triplet. We also show
other regression versions in the plot. The second regression is
computed without the triplet members mentioned above: it is shown
by the red solid line with the slope of \mbox{$k2 =1.01\pm0.08$} (52
objects). The third regression is based on the galaxies within the
narrow luminosity range of \mbox{$\log L_{B}= 7.0$--$8.6$} and
shown by the red dash-dotted line with a slope of \mbox{$k3
=1.21\pm0.12$} (45~objects).  A comparison of the slope difference
with its ``combined error'' indicates that the slopes differ at
the level of $2.5$--$3\sigma$: \mbox{$k1-k2=-0.17\pm0.056$},
\mbox{$k1-k3=-0.35\pm0.144$}. We can thus conclude that the nature
of the relation between the two global parameters of low-mass void
galaxies over a wide luminosity range is so far not entirely
clear. Extensive samples are needed, especially for the
\mbox{$\log L_{B} < 7.5$} domain. Extending the analysis to other
galaxies of this void and to low-mass galaxies from other nearby
voids will increase the total number of galaxies involved by a
factor of  \mbox{2--3} and provide a better understanding of the
discussed relation. 
\begin{figure}
\setcaptionmargin{5mm} \onelinecaptionsfalse
\captionstyle{normal}
 \includegraphics[angle=-90,width=\columnwidth,bb=0 25 539 765,clip]{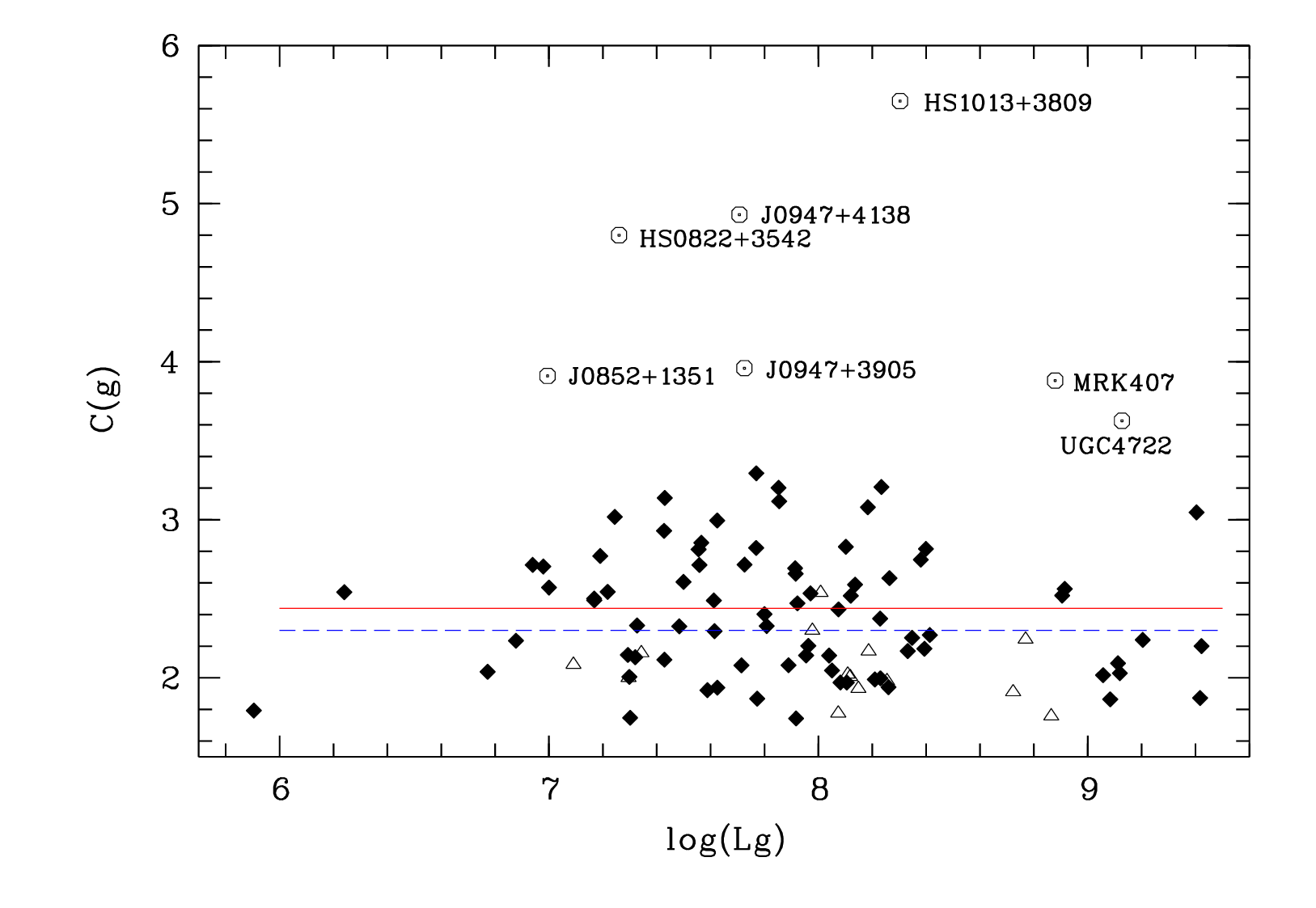}
\caption{\label{fig:logLg_R90_R50} Relation
between the luminosity $\log L_{g}$ and the $g$-band concentration
index $C(g)$. The red solid line shows the mean value ($R=2.44$)
averaged over all the objects of our sample (the filled diamonds).
The blue dashed line shows the mean value ($R=2.3$) from Garcia-Appadoo
et al.~\cite{Garcia09} for their sample of 195
objects. The open triangles show the group of the so-called
``inchoate''\ galaxies pointed out by the above authors. }
\end{figure}
\begin{figure}
\setcaptionmargin{5mm} \onelinecaptionsfalse
\captionstyle{normal}
 \includegraphics[angle=-90,width=0.98\columnwidth]{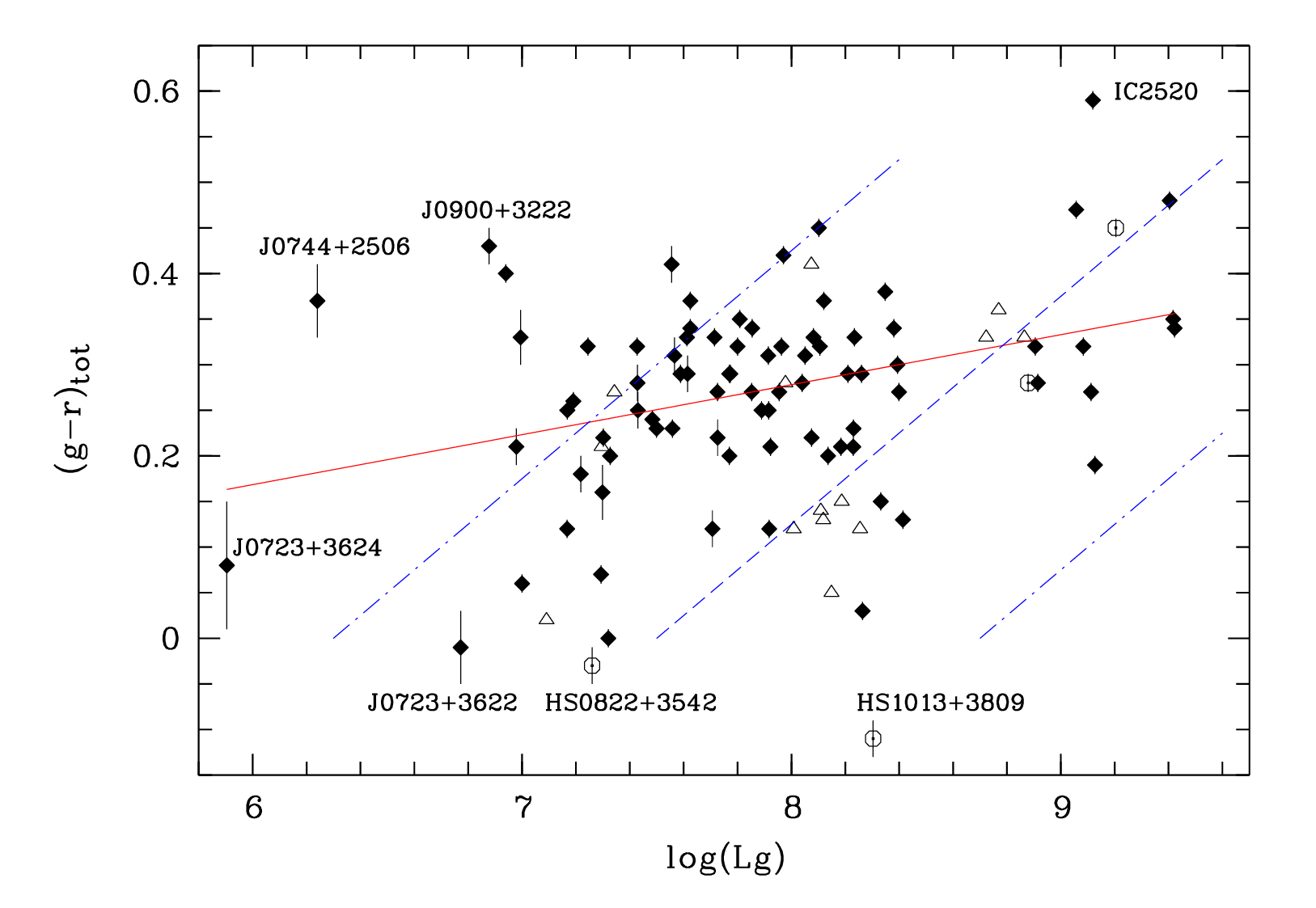}
 \vspace{3pt}
\caption{\label{fig:logLg_g_r_tot} Relation
between the $g$-band luminosity $\log L_{g}$ and integrated color
$(g-r)_{\rm tot}$. The red solid line shows the linear regression
based on all the objects of our sample (the filled diamonds)
except for the BGCs (the open circles). The blue dashed line shows the
linear regression from~Garcia-Appadoo et
al.~\cite{Garcia09}, the blue dash-dotted lines show
the scatter of the data points of their sample, and the open
triangles show the group of ``inchoate''\ galaxies pointed out by
the above authors. }
\end{figure}

\begin{figure}
\setcaptionmargin{5mm} \onelinecaptionsfalse
\captionstyle{normal}
 \includegraphics[angle=-90,width=\columnwidth]{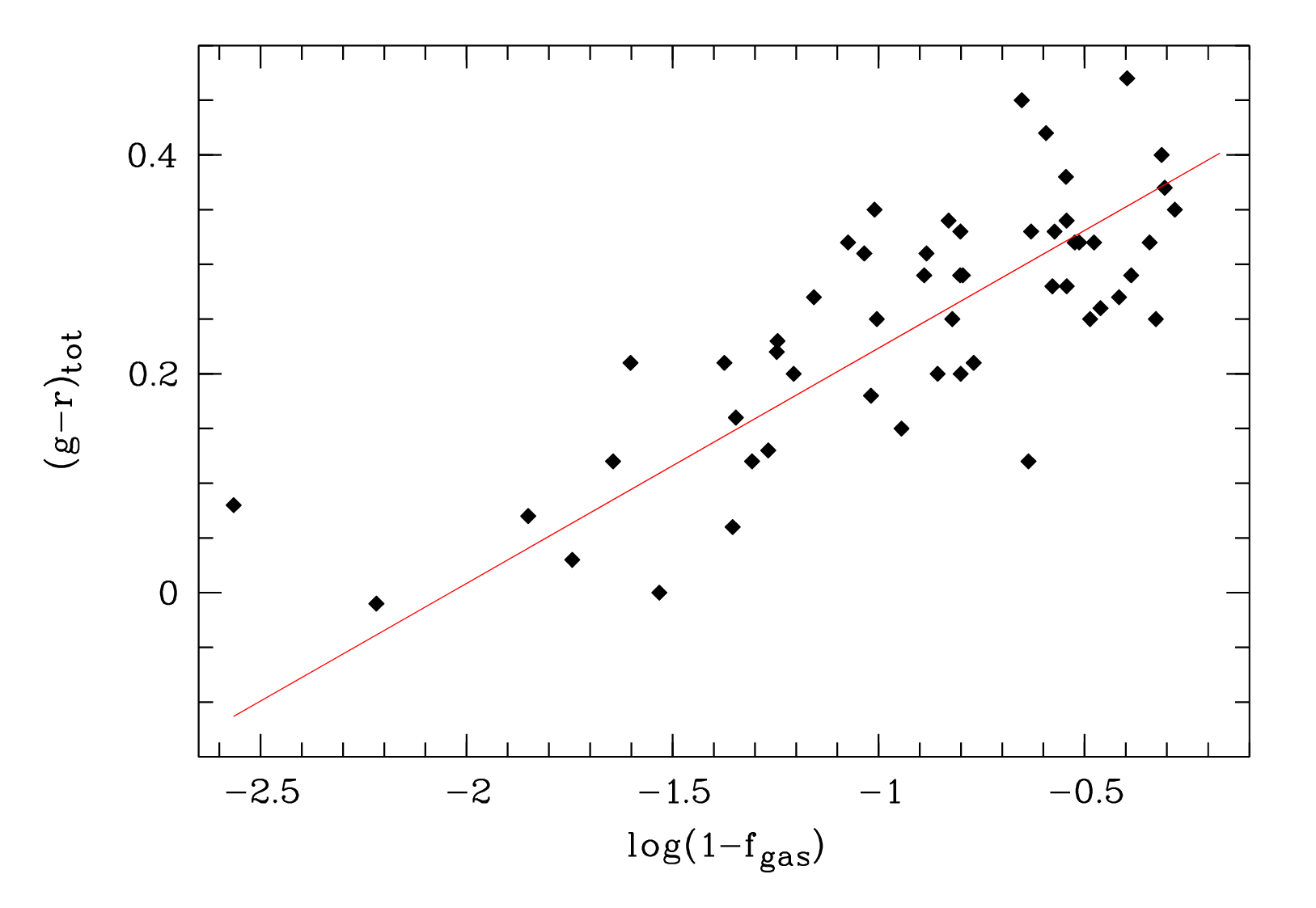}
\caption{\label{fig:f_gas_g-r_t} Relation between
the stellar mass fraction (\mbox{$\log(1-f_{\rm gas})$}) and
integrated color $(g-r)_{\rm tot}$. The red solid line shows the
linear regression based on the objects of our sample. }
\end{figure}

As is evident from Fig.~\ref{fig:logLg_R90_R50},
the  $g$-band concentration index $C(g)=R_{\rm 90g}/R_{\rm 50g}$
(where $R_{\rm 90}(g)$ and $R_{\rm 50}(g)$ are the radii
containing  90\% and half of the  \mbox{$g$-band} flux of the
galaxy) does not depend on luminosity~$L_{g}$. The mean
concentration index averaged over the galaxies of our sample is
$C(g)=2.44$ (the red solid line) which is close to the value $2.32$
for purely exponential disks. When computing our mean value, we
excluded, like in the above cases, the BCGs, very compact objects
(J0947+4138, J0947+3905, and J0852+1351), and the perturbed disk
galaxy UGC~4722 with a tidal tale and an intense burst of star
formation (all these objects are shown by open circles). We also
show the mean concentration $C(g)=2.3$ for the galaxies from
Garcia-Appadoo et al.~\cite{Garcia09} (the blue dashed
line) and the positions of their ``inchoate''\ galaxies (the open
triangles). 

Figure~\ref{fig:logLg_g_r_tot} shows the
relationship between the $g$-band luminosity logarithm $\log
L_{g}$ and the integrated color \mbox{$(g-r)_{\rm tot}$} for void
galaxies. The red solid line shows the linear regression for this
sample \mbox{($k =0.05\pm0.02$)}. For comparison, we show by the
blue dashed line the linear regression for the galaxies of the  ES
survey ($k =0.25$) and the positions of the  ``inchoate''\
galaxies from~\cite{Garcia09} (the open
triangles). Whereas the galaxies of the  ES sample exhibit a
strong trend, the $(g-r)_{\rm tot}$ colors (like the parameter
$\mu_{\rm eff}(g)$  in
Fig.~\ref{fig:logLg_mu_eff}) on the average vary
only slightly with decreasing luminosity: from $0.35$ to about
$0.15$, albeit with a large scatter. To understand the nature of
this scatter better, galaxies with extreme  $(g-r)_{\rm tot}$
parameters should be studied in more detail.

Figure~\ref{fig:f_gas_g-r_t} shows the
relationship between the stellar mass fraction ($\log(1-f_{\rm
gas})$) and integrated color $(g-r)_{\rm tot}$. The red solid line
shows the linear regression between these parameters \mbox{($k
=0.215\pm0.024$)}. As expected, in this case the relation shows up
more conspicuously than in the previous figure, because the
coefficient $\Upsilon$ used to estimate the stellar mass from
luminosity depends on the $(g-i)$ color (which correlates with
$g-r$)---see Section~3.2. The bluer the color, the smaller the
stellar mass, and, for a fixed gas mass, the smaller the stellar
mass fraction.

\begin{figure}
\setcaptionmargin{5mm} \onelinecaptionsfalse
\captionstyle{normal}
 \includegraphics[angle=-90,width=\columnwidth]{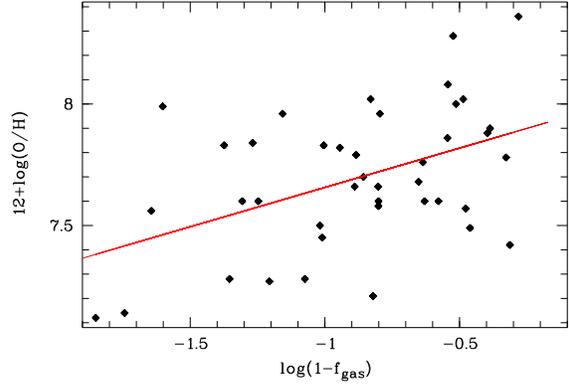}
\caption{\label{fig:f_gas_O_H} Relation between
the stellar mass fraction (\mbox{$\log(1-f_{\rm gas})$}) and
metallicity $12+\log({\rm O/H})$. The red solid line shows the linear
regression based on the objects of our sample. }
\end{figure}

Figure~\ref{fig:f_gas_O_H} shows the relationship
between the stellar mass fraction ($\log(1-f_{\rm gas})$) and the
oxygen abundance in the interstellar medium  (similar to
metallicity), $12+\log({\rm O/H})$, for the void galaxies for
which this parameter has been measured. Like in the previous
figures, the red solid line shows the linear regression
($k=0.32\pm0.10$). The apparent trend of the decrease of
metallicity with decreasing stellar mass fraction is consistent
with what is to be expected in the so-called  ``closed box'' model
of galaxy evolution (i.e., evolution without exchange with the
surrounding medium). However, the scatter of the parameter ${\rm
O/H}$ is so large that the ``closed box'' approximation is often
invalid in void galaxies, including  LSBD galaxies.

\subsection{Two-Color Diagrams and Age Estimates}

\begin{table*}
\setcaptionmargin{0mm} \onelinecaptionstrue \captionstyle{normal}
{\bf Table 5.} Parameters of peculiar galaxies in the void
\medskip
\begin{tabular}{l|c|c|c|c|c|c|c|c}
\hline
Parameter        & J0723+3622&J0723+3624 & J0737+47& J0812+48 & J0926+33 & J0929+25  & S\,0822+3545  & U\,3966      \\
\hline
$T_{\rm old}$, Gyr & 1$-$2 & 2$-$2.5  & 2$-$3 & 4$-$6    & 3$-$3.5  & 3$-$5   & 1$-$3    & 2.5$-$3     \\
$f_{\rm gas}$      & 0.997 & 0.997    & 0.96  & 0.94     & 0.99     & 0.95    & 0.97     & 0.98       \\
12+$\log({\rm O/H})$ & $-$ & $-$      & 7.28  & 7.27     & 7.12     & 7.2:    & $-$     & 7.56       \\
$M_{B}$            &$-$11.79&$-$9.57  &$-$12.54& $-$13.08& $-$12.91 &$-$12.95 & $-$13.11 & $-$14.58    \\
\hline
\end{tabular}
\end{table*}

In
Figs.~\ref{fig:ugr_periph}~and~\ref{fig:gri_periph}
we compare the colors of the outer regions of 81 void galaxies
with {\tt PEGASE2} models for two extreme cases of the
star-formation (SF) law: instantaneous ($\rm inst$) SF and
continuous ($\rm const$) SF with a constant rate. We adopt two
initial stellar mass functions (IMF): the standard  Salpeter IMF
and the IMF of Kroupa. For the sake of illustration we show the
tracks for the metallicity of $Z=0.002$, which lies approximately
in the middle of the range of known metallicities of void
galaxies. The contribution of nebular emission may be significant
for the four BCGs in our sample. We excluded these objects from
further analysis because the estimates of their stellar population
colors are uncertain.

The colors of the outer regions of 77 of the 81~void galaxies
studied here agree rather well with the {\tt PEGASE2} evolutionary
tracks for the case of continuous star formation. The time $T_{\rm
SF}$ elapsed since the onset of the star-formation (or in
other terms, the age of a galaxy) spans a wide range,
approximately from 1 to 14~Gyr. Most of these galaxies better fit
the tracks computed with the Kroupa IMF. The colors of the outer
parts of the majority of the objects correspond to typical
galactic ages: \mbox{$T\sim10$--$13$~Gyr}. However, the colors of
the outer regions of seven objects correspond to the times $T_{\rm
SF}$ as small as about  $1$--$3.5$~Gyr. The colors of \mbox{six}
other galaxies correspond to intermediate times \mbox{$T_{\rm SF}
\sim4$--$6.5$~Gyr}. We discuss these galaxies and their
peculiarities in more detail in the next  section.

Only for four galaxies---J0744+2506, IC\,2450, J0928+2845, and
CGCG\,035-007---their rather red \linebreak \mbox{($g-r
\sim0.5$--$0.6$)} colors, unlike those of the remaining 77
galaxies, can be interpreted as a result of a rather short
(``instantaneous'') and very old (about 10~Gyr) star formation
episode, typical of elliptical galaxies. However, their morphology
is inconsistent with such a hypothesis. The presence of emission
regions either near the center, like in  J0744+2506 and IC\,2450,
or shifted towards the edge, like in J0928+284 and CGCG\,035-007,
is indicative of recent star formation caused either by external
perturbation or by the infall of fresh gas. These galaxies deserve
a more thorough investigation as objects whose outer regions have
colors atypical of the sample studied. In particular, it is
important to understand what fraction of the fainter of such
objects remains undiscovered because of observational selection if
the star formation in them has not been triggered by any external
factors over the past several tens of Myr.

\subsection{Peculiar Galaxies of the Void}

The peculiar properties of some galaxies with non-cosmological
times elapsed since the onset of the main star formation episode
(SAO\,0822+3542, UGC\,5340~$=$~DDO\,68, J0926+3343, J0723+3622,
J0723+3624,  and J0737+4724) were already pointed out in the
papers dedicated to the individual galaxies of this
void~\cite{Pustilnik03,DDO68,DDO68_sdss,J0926,PaperIII,Triplet}.

Here we give for these galaxies an independent corroboration of
the blue colors of their peripheral regions which corresponds to
short times elapsed since the onset of continuous star formation:
$T_{\rm SF} \lesssim$ 3.5~Gyr. For two other galaxies---UGC\,3966,
UGC\,4117---such data have been obtained for the first time. The
colors of the outer regions of UGC\,3672, UGC\,3860, J0812+4836,
J0929+2502, UGC\,5272 and its companion\linebreak UGC\,5272b also
correspond to non-cosmological times in the $T_{\rm SF} range of
\sim3$--$6.5$~Gyr.

Table~5 summarizes the evolutionary
parameters for eight galaxies which can be classified as unevolved
based on the combination of their properties. Their gas mass
fractions are equal to \mbox{94--99.7\%}, and their metallicities
$12+\log({\rm O/H})$ are 2-5 times lower than those of typical
irregular dwarf galaxies of the same luminosity but residing in
environments with intermediate or high density of galaxies
(see~\mbox{\cite{Pustilnik03,
SBS0335BTA, DDO68,
DDO68_sdss, J0926}}), whereas the
colors of the visible old stellar population in these objects
correspond to the epoch of the onset of star formation $T_{\rm SF}
\lesssim 1$--$5$~Gyr. We place the object J0723+3624 with
$M_{B}=-9.57$ in the same group of galaxies. The  UGC\,4117 galaxy
has blue peripheral colors corresponding to  $T_{\rm SF}
\sim2$~Gyr but does not fit into the list of peculiar galaxies of
the void because its metallicity \mbox{$12+\log({\rm O/H})=7.82$}
does not differ very much from that expected for its luminosity
(\mbox{$M_{B}=-15.6$}). The same is true for the galaxies
UGC\,3672, UGC\,3860, UGC\,5272, and UGC\,5272b with peripheral
ages ranging from $5$~to~$6$~Gyr. New Hubble Space Telescope data
for UGC\,5340 (DDO\,68)~\cite{Tikhonov14} confirm
the suggestions of Pustilnik et al.~\cite{DDO68}
and Ekta et al.~\cite{ekta08} that this galaxy is
made up of two merging components with very different properties
and that this fact should be taken into account when associating
the object with the group of peculiar galaxies. The more massive
central component with $M_{B} \sim -16$ has an old stellar
population with the metallicity five times lower than solar and
hence is a rather typical late-type galaxy. The stars in the much
less massive component (UGC\,5340b), extending along the eastern
edge of the more massive component, have metallicities no greater
than $Z_{\odot}/20$ which is consistent with the gas metallicity
\mbox{$12+\log({\rm O/H})=7.14$} estimated for the H\,II regions
of the galaxy. The ages of most or all of its stars do not exceed
$2$~Gyr. However, thus far the mass or luminosity of the less
massive component cannot be estimated.

A closer analysis of the properties of the most peculiar galaxies
mentioned above suggests that their fraction is much higher among
less luminous galaxies. However, this may be an accidental result
due to the statistically small number of bright void galaxies. The
hypothesis about a relation between the luminosity and the
fraction of unevolved galaxies in the void can be tested using the
statistical criterion ``2$\times$2~contingency
table''~\cite{Bolshev83}), which is widely used in
biology and applied studies. A detailed description of this
criterion can be found in the paper by Pustilnik et
al.~\cite{Pustilnik95}, who used it for
astronomical applications.

The idea of the method consists of analyzing the numbers in a
2$\times$2 contingency table, which correspond to different
combinations of two properties of the elements of the sample. In
our case we adopt the property $Y$ of being a low-luminosity
galaxy with $M_{B} > M_{\rm faint}$, where  $M_{\rm faint}$ is the
threshold value based on certain considerations. Thus, the
property $Y$ breaks the sample into the $Y$ and non-$Y$
subsamples. Similarly, the second property $Z$ is that of being an
unevolved galaxy according to the above criteria. Correspondingly,
\mbox{non-$Z$} means that the object is a galaxy with a more
standard visible stellar population. To test the zero hypothesis
$h_{\mathrm 0}$ that the two properties in the sample considered
are independent of each other, we must compose a 2$\times$2 table
in the following form:

\vspace{1mm} \hspace{-9mm} \centerline{
\begin{tabular}{l@{~~~~}c@{~~~~~}c@{~~~~~}c} 
Property & $Y$     & non-$Y$      & Total \\  
$Z$      & $m$     & $n-m$        & $n$   \\[-5pt]
non-$Z$  & $M-m$   & $N-n-(M-m)$  & $N-n$ \\[-5pt]  
Total    & $M$     & $N-M$        &  $N$  \\
\end{tabular}
} \vspace{1mm}

\noindent Here $m$, $n-m$,  $M-m$,  and $N-n-(M-m)$ are the
numbers of galaxies in the sample having the property combinations
($Y$,\,$Z$), (non-$Y$,\,$Z$), ($Y$,\,\mbox{non-$Z$}),  and
(non-$Y$,\,non-$Z$) respectively. As is shown in the
book~\cite{Bolshev83} (pp.~77-78), if the
properties  $Y$ and $Z$ are independent of each other, the
probability of obtaining the contingency table with such numbers
is described by a hypergeometric distribution, which at $N > 25$
can be well approximated by the so-called incomplete beta function
$I_x (a,b)$, where the parameters $x$,~$a$,~and~$b$ can be
expressed in terms of the numbers in the 2$\times$2 contingency
table by the formulas (27)-(30) on p.~74 of the
book~\cite{Bolshev83} (see also Appendix
to~\cite{Pustilnik95}).

If there is no real correlation in the sample of void galaxies
between the properties of low luminosity and ``small'' age, then
in the case of the observed distribution of absolute magnitudes
and whatever threshold $M_{\rm faint}$ is adopted, the numbers in
the contingency table should correspond to a relatively low
probability of rejecting the null hypothesis.

To account more fully for all the available information on the
unevolved galaxies of the Lynx-Cancer void sample, we considered
the object J0723+3624 with the absolute magnitude \mbox{$M_{B}
=-9.57$} to be an unevolved object when analyzing the 2$\times$2
contingency table. Despite the failure to measure the peripheral
colors of this very small galaxy, which is a member of the unusual
triplet in the central part of the void, the record high gas mass
fraction of this object (0.997) and its blue integrated color
result in age estimates of less than \mbox{$2$--$3$~Gyr} in any
reasonable evolutionary scenario (Fig.~7
in~\cite{Triplet}). In view of the above, the
\mbox{2$\times$2}~contingency table for \mbox{$M_{\rm
faint}=-13.15$} is:

\vspace{1mm} \centerline{
\begin{tabular}{c@{~~~~~}c@{~~~~~}c@{~~~~~}c} 
Property    & $Y$ & non-$Y$      & Total  \\   
$Z$        & 6   & 12          & 18     \\[-5pt]
non-$Z$     & 2   & 65          & 67     \\[-5pt]   
Total      & 8   & 77          & 85     \\
\end{tabular}
} \vspace{1mm}

\noindent The probability of the table with these values computed
by the appropriate formulas for the incomplete beta function
yields a statistically significant result: the probability of
rejecting the null hypothesis is \mbox{$P=0.9993$}. This result
should be interpreted as the existence of a significant
statistical relation between the two properties in the sense that
the fraction of unevolved galaxies is significantly higher in the
group with the absolute magnitudes fainter than \mbox{$M_{\rm
faint}=-13.15$} (a simple comparison of the fractions  6/18 and
2/67 indicates a difference of more than one order of magnitude).
For this computation we included the unevolved component
UGC\,5340b in the sample, assuming that its  $M_{B}$ is brighter
than~$-13.15$. Otherwise the significance level would be even
higher.

The discovery of such a relation has two important implications.
The first one is associated with the understanding of the physics
of the processes resulting in unevolved galaxies appearing only
among sufficiently low-mass objects. The second implication is a
methodological one. The correlation found indicates that to search
effectively for such objects in the voids, one must study galaxies
with absolute magnitudes fainter than or near the ``threshold.\!''

Let us return to the situation with the blue colors of the
peripheral regions of some galaxies, where the parameter ${\rm
O/H}$ is only slightly lower than expected for their luminosity
(UGC\,3672, 3860, 4117, 5272, and 5272b). These objects do not
resemble unevolved galaxies, but their unusual colors require
further discussion. They can be explained by a relatively recent
perturbation (on a time scale of \mbox{2--6~Gyr}) which resulted
in an increased rate of star formation, the traces of which are
now also visible in their outer regions. The optical morphology
of the UGC\,3672, UGC\,3860, and UGC\,4117 galaxies and that of
the pair UGC\,5272/5272b is indeed indicative of a significant
perturbation and/or ongoing interaction. Further H\,I mapping of
these objects should provide more compelling evidence for their
perturbed state.

\section{CONCLUSIONS}
\label{sec:summ}

To summarize our study of void galaxies including their
statistical analysis and a comparison with other data and samples,
we formulate the following results and conclusions.

\begin{list}{}{
\setlength\leftmargin{2mm} \setlength\topsep{2mm}
\setlength\parsep{0mm} \setlength\itemsep{2mm}}

\item (1) We determined the photometric parameters in the
$u,g,r,i$ bands (integrated magnitudes and colors, effective radii
and the corresponding surface brightness values, the optical and
Holmberg radii) for 85~galaxies of the Lynx-Cancer void based on
their images adopted from the SDSS database. We analyzed the
radial surface brightness profiles to determine the central
surface brightness levels and scale lengths of the model disks.

\item (2) We compared the  $(u-g), (g-r), (r-i)$ color indices of
the outer regions of the galaxies outside the regions of recent
star formation with the {\tt PEGASE2} evolutionary tracks and
estimated the time~$T_{\rm SF}$ elapsed since the onset of the
star-formation. For about 85\% of the galaxies, these time
scales lie in the range of \mbox{$T \sim10$--$13$~Gyr}, which is
typical of galaxies in denser environments. The colors of 13
galaxies imply much smaller  $T_{\rm SF}$. The corresponding times
for seven galaxies lie in the range of  $1$--$3.5$~Gyr, and those
of the six remaining galaxies range approximately from $4$ to
$6.5$~Gyr.

\item (3) Most of the galaxies with unusual colors of
outer regions have very low gas metallicities (\mbox{$12+\log({\rm
O/H})\lesssim7.4$} which are 2-5 times lower than expected for
their luminosity) and the highest values of the empirical gas
content parameter \mbox{$\M$(HI)$/L_{B} \gtrsim 2.5$}. We
estimated the stellar masses of the galaxies of this group and
found their gas mass fractions to be in the range of 94-99\%
which is the highest value of this parameter ever determined. By
their properties these galaxies are unevolved.

\item (4) A comparison of the statistical relations between the 
observed properties of the
galaxies in the Lynx-Cancer void and galaxies from the Equatorial 
Survey (ES) initially
selected by their H\,I line emission shows the similarity of the 
two samples throughout the common
luminosity range. However, void galaxies contain a significant
fraction of objects that
do not follow the common trends and relations. A small group of 
the so-called ``inchoate''\
galaxies in the ES survey exhibits unusual properties. More detailed 
studies will probably show
what part of them are analogs of unusual galaxies in the voids.

\item (5) The group of unevolved galaxies accounts for about 15\%
of all the low surface brightness dwarf galaxies in the void and
is represented substantially more often among low luminosity
galaxies (\mbox{$M_{B} >-13.2$}), where their fraction amounts to
about 30\%. This fact confirms that the environment has a stronger
effect on less massive galaxies and provides a clue to an
efficient search for such unusual galaxies.
\end{list}

\begin{acknowledgements}

We are grateful to D.~I.~Makarov for his numerous useful critical
comments and suggestions, which helped improve the paper
substantially. Yu.~A.~Perepelitsyna and S.~A.~Pustilnik
acknowledge the support of the Russian Foundation for Basic
Research (grants No.~11-02-00261, \mbox{13-02-90407},
and\linebreak \mbox{13-02-90734}) and partial support of the
Federal target programme {\it Research and Pedagogical\linebreak
Cadre for Innovative Russia} (proposal\linebreak
No.~\mbox{2012-1.5-12-000-1011-004}, agreement number~8523).
A.~Yu.~Kniazev acknowledges the support of the National Research
Foundation (NRF) of South Africa. We are grateful to the SDSS
consortium for the spectroscopic, photometric, and auxiliary
information on the studied galaxies available from the SDSS
database~\cite{DR7}.
\end{acknowledgements}

\onecolumngrid


\clearpage
\landscape{
\clearpage
\renewcommand{\baselinestretch}{1.0}
\begin{table*}

\setcaptionmargin{0mm} \onelinecaptionstrue
\captionstyle{flushleft}

\begin{center}
\vspace{-0.5cm}
\hspace{2.5cm} 
{\bf Table 1.}\ Basic parameters of the galaxies in the Lynx?Cancer void.

\vspace{0.3cm}
\begin{tabular}{r|l|c|l|r|r|r|r|r|r|r|r|r|r|c|r|} \hline \\[-0.2cm]
\multicolumn{1}{c|}{}       &
\multicolumn{1}{c|}{Name or}   &
\multicolumn{1}{c|}{Name in brief} &
\multicolumn{1}{c|}{Type}       &
\multicolumn{1}{c|}{$V_{\rm hel}$} &
\multicolumn{1}{c|}{$Dist$}  &
\multicolumn{1}{c|}{$A_{\rm B}^{\rm NED}$}  &
\multicolumn{1}{c|}{$B_{\rm tot}$}  &
\multicolumn{1}{c|}{$M_{\rm B,0}$}  &
\multicolumn{1}{c|}{$B_{\rm tot}^{\rm NED}$}  &
\multicolumn{1}{l|}{$\M$(\HI)} &
\multicolumn{1}{c|}{$\M$(\HI)/$L_{\rm B}$} &
\multicolumn{1}{l|}{$\M_{\rm star}$}      &
\multicolumn{1}{l|}{$f_{\rm gas}$}      &
\multicolumn{1}{c|}{{\tiny 12+log(O/H)}}      &
\multicolumn{1}{l|}{b/a}       \\
&
\multicolumn{1}{c|}{prefix }  &
\multicolumn{1}{c|}{IAU format }  &
\multicolumn{1}{l|}{}     &
\multicolumn{1}{c|}{{\tiny km/s}}  &
\multicolumn{1}{r|}{{\tiny Mpc}}  &
\multicolumn{1}{r|}{{\tiny mag}}  &
\multicolumn{1}{r|}{{\tiny mag}}  &
\multicolumn{1}{r|}{{\tiny mag}}  &
\multicolumn{1}{r|}{{\tiny mag}}  &
\multicolumn{1}{c|}{{\tiny$10^7M_{\odot}$}}  &
\multicolumn{1}{c|}{}  &
\multicolumn{1}{l|}{{\tiny$10^7M_{\odot}$}}  &
\multicolumn{1}{l|}{ }     &
\multicolumn{1}{l|}{ }      &
\multicolumn{1}{l|}{ }     \\
&
\multicolumn{1}{c|}{ (1) }  &
\multicolumn{1}{c|}{ (2) }  &
\multicolumn{1}{c|}{ (3) }  &
\multicolumn{1}{c|}{ (4) }  &
\multicolumn{1}{c|}{ (5) }  &
\multicolumn{1}{c|}{ (6) }  &
\multicolumn{1}{c|}{ (7) }  &
\multicolumn{1}{c|}{ (8) }  &
\multicolumn{1}{c|}{ (9) }  &
\multicolumn{1}{c|}{ (10) } &
\multicolumn{1}{c|}{ (11) } &
\multicolumn{1}{c|}{ (12) } &
\multicolumn{1}{c|}{ (13) } &
\multicolumn{1}{c|}{ (14) } &
\multicolumn{1}{c|}{ (15) } \\
\\[-0.2cm] \hline \\[-0.2cm]
 \qq& HIPASSJ0626+24   & J0626+2439 & Scd &1473&23.04&1.79& 17.98 & -15.62 & 17.60 &  --   &--    &  5.53&--  &--    & 0.59 \\
\end{tabular}
 \end{center}
 \end{table*}
 
 \newpage
\clearpage 
\renewcommand{\baselinestretch}{1.0}
\begin{table*}

\setcaptionmargin{0mm} \onelinecaptionstrue
\captionstyle{flushleft}

\begin{center}
\vspace{-0.5cm}
\hspace{2.5cm}
{\bf Table 1.}\ Basic parameters of the galaxies in the Lynx?Cancer void. (Condt.)

\vspace{0.3cm}
\begin{tabular}{r|l|c|l|r|r|r|r|r|r|r|r|r|r|c|r|} \hline \\[-0.2cm]
\multicolumn{1}{c|}{}       &
\multicolumn{1}{c|}{Name or}   &
\multicolumn{1}{c|}{Name in brief} &
\multicolumn{1}{c|}{Type}       &
\multicolumn{1}{c|}{$V_{\rm hel}$} &
\multicolumn{1}{c|}{$Dist$}  &
\multicolumn{1}{c|}{$A_{\rm B}^{\rm NED}$}  &
\multicolumn{1}{c|}{$B_{\rm tot}$}  &
\multicolumn{1}{c|}{$M_{\rm B,0}$}  &
\multicolumn{1}{c|}{$B_{\rm tot}^{\rm NED}$}  &
\multicolumn{1}{l|}{$\M$(\HI)} &
\multicolumn{1}{c|}{$\M$(\HI)/$L_{\rm B}$} &
\multicolumn{1}{l|}{$\M_{\rm star}$}      &
\multicolumn{1}{l|}{$f_{\rm gas}$}      &
\multicolumn{1}{c|}{{\tiny 12+log(O/H)}}      &
\multicolumn{1}{l|}{b/a}       \\
&
\multicolumn{1}{c|}{prefix }  &
\multicolumn{1}{c|}{IAU format }  &
\multicolumn{1}{l|}{}     &
\multicolumn{1}{c|}{{\tiny km/s}}  &
\multicolumn{1}{r|}{{\tiny Mpc}}  &
\multicolumn{1}{r|}{{\tiny mag}}  &
\multicolumn{1}{r|}{{\tiny mag}}  &
\multicolumn{1}{r|}{{\tiny mag}}  &
\multicolumn{1}{r|}{{\tiny mag}}  &
\multicolumn{1}{c|}{{\tiny$10^7M_{\odot}$}}  &
\multicolumn{1}{c|}{}  &
\multicolumn{1}{l|}{{\tiny$10^7M_{\odot}$}}  &
\multicolumn{1}{l|}{ }     &
\multicolumn{1}{l|}{ }      &
\multicolumn{1}{l|}{ }     \\
&
\multicolumn{1}{c|}{ (1) }  &
\multicolumn{1}{c|}{ (2) }  &
\multicolumn{1}{c|}{ (3) }  &
\multicolumn{1}{c|}{ (4) }  &
\multicolumn{1}{c|}{ (5) }  &
\multicolumn{1}{c|}{ (6) }  &
\multicolumn{1}{c|}{ (7) }  &
\multicolumn{1}{c|}{ (8) }  &
\multicolumn{1}{c|}{ (9) }  &
\multicolumn{1}{c|}{ (10) } &
\multicolumn{1}{c|}{ (11) } &
\multicolumn{1}{c|}{ (12) } &
\multicolumn{1}{c|}{ (13) } &
\multicolumn{1}{c|}{ (14) } &
\multicolumn{1}{c|}{ (15) } \\
\\[-0.2cm] \hline \\[-0.2cm]
 \qq& SDSS             & J0831+4104 & LSB &582 &11.64&0.16& 17.71 & -12.78 & 17.71 &  0.58 & 0.29 &  0.59&0.57&--    & 0.90 \\
\end{tabular}
  \end{center}
  \end{table*}
 
 \newpage 
 \renewcommand{\baselinestretch}{1.0}
 \begin{table*}

\setcaptionmargin{0mm} \onelinecaptionstrue
\captionstyle{flushleft}

\begin{center}
\vspace{-0.5cm}
\hspace{2.5cm}
{\bf Table 1.}\ Basic parameters of the galaxies in the Lynx?Cancer void. (Condt.)

\vspace{0.3cm}

\begin{tabular}{r|l|c|l|r|r|r|r|r|r|r|r|r|r|c|r|} \hline \\[-0.2cm]
\multicolumn{1}{c|}{}       &
\multicolumn{1}{c|}{Name or}   &
\multicolumn{1}{c|}{Name in brief} &
\multicolumn{1}{c|}{Type}       &
\multicolumn{1}{c|}{$V_{\rm hel}$} &
\multicolumn{1}{c|}{$Dist$}  &
\multicolumn{1}{c|}{$A_{\rm B}^{\rm NED}$}  &
\multicolumn{1}{c|}{$B_{\rm tot}$}  &
\multicolumn{1}{c|}{$M_{\rm B,0}$}  &
\multicolumn{1}{c|}{$B_{\rm tot}^{\rm NED}$}  &
\multicolumn{1}{l|}{$\M$(\HI)} &
\multicolumn{1}{c|}{$\M$(\HI)/$L_{\rm B}$} &
\multicolumn{1}{l|}{$\M_{\rm star}$}      &
\multicolumn{1}{l|}{$f_{\rm gas}$}      &
\multicolumn{1}{c|}{{\tiny 12+log(O/H)}}      &
\multicolumn{1}{l|}{b/a}       \\
&
\multicolumn{1}{c|}{prefix }  &
\multicolumn{1}{c|}{IAU format }  &
\multicolumn{1}{l|}{}     &
\multicolumn{1}{c|}{{\tiny km/s}}  &
\multicolumn{1}{r|}{{\tiny Mpc}}  &
\multicolumn{1}{r|}{{\tiny mag}}  &
\multicolumn{1}{r|}{{\tiny mag}}  &
\multicolumn{1}{r|}{{\tiny mag}}  &
\multicolumn{1}{r|}{{\tiny mag}}  &
\multicolumn{1}{c|}{{\tiny$10^7M_{\odot}$}}  &
\multicolumn{1}{c|}{}  &
\multicolumn{1}{l|}{{\tiny$10^7M_{\odot}$}}  &
\multicolumn{1}{l|}{ }     &
\multicolumn{1}{l|}{ }      &
\multicolumn{1}{l|}{ }     \\
&
\multicolumn{1}{c|}{ (1) }  &
\multicolumn{1}{c|}{ (2) }  &
\multicolumn{1}{c|}{ (3) }  &
\multicolumn{1}{c|}{ (4) }  &
\multicolumn{1}{c|}{ (5) }  &
\multicolumn{1}{c|}{ (6) }  &
\multicolumn{1}{c|}{ (7) }  &
\multicolumn{1}{c|}{ (8) }  &
\multicolumn{1}{c|}{ (9) }  &
\multicolumn{1}{c|}{ (10) } &
\multicolumn{1}{c|}{ (11) } &
\multicolumn{1}{c|}{ (12) } &
\multicolumn{1}{c|}{ (13) } &
\multicolumn{1}{c|}{ (14) } &
\multicolumn{1}{c|}{ (15) } \\
\\[-0.2cm] \hline \\[-0.2cm]
 \qq& UGC5272b         & J0950+3127 & Im   &539 &10.27&0.10& 17.56 & -12.60 & 17.68 &  2.89 & 1.70 &  0.19&0.95& 7.60& 0.67 \\
\end{tabular}
 \end{center}
  \end{table*}

\clearpage
\clearpage
\headsep 1cm
\renewcommand{\baselinestretch}{1.0}

\setcounter{qub}{0}
\hoffset=0cm
\begin{table*}
\setcaptionmargin{0mm} \onelinecaptionstrue
\captionstyle{flushleft}
                                                               
\begin{center}
\vspace{-0.5cm}
\rotatebox{-90}{%
\parbox{\textwidth}{%
{\bf Table 2.} Photometric paramerets I. \\
\vspace{0.3cm}
\begin{tabular}{r|l|r|r|r|r|r|r|r|r|c|c|c} \hline \\[-0.2cm]
\multicolumn{1}{c|}{} &
\multicolumn{1}{c|}{Name} &
\multicolumn{1}{c|}{$R_{\rm 50}(g)$}   &
\multicolumn{1}{c|}{$R_{\rm 90}(g)$}  &
\multicolumn{1}{c|}{$\mu_{\rm eff}(g)$} &
\multicolumn{1}{c|}{$R^{25}(g)$} &
\multicolumn{1}{c|}{$a_{\rm 25}$} &
\multicolumn{1}{c|}{$g^{25}$} &
\multicolumn{1}{c|}{$R^{26.5}(g)$} &
\multicolumn{1}{c|}{$g^{26.5}$} &
\multicolumn{1}{c|}{Opt.size} &
\multicolumn{1}{c|}{Opt.size} \\
&
\multicolumn{1}{c|}{}  &
\multicolumn{1}{c|}{$''$}  &
\multicolumn{1}{c|}{$''$}  &
\multicolumn{1}{c|}{$^{m}/\square''$}  &
\multicolumn{1}{c|}{$''$}  &
\multicolumn{1}{c|}{\tiny kpc}  &
\multicolumn{1}{c|}{\tiny mag}  &
\multicolumn{1}{c|}{$''$}  &
\multicolumn{1}{c|}{\tiny mag}  &
\multicolumn{1}{c|}{$''$}  &
\multicolumn{1}{c|}{\tiny kpc}  \\
&
\multicolumn{1}{c|}{ (1) }  &
\multicolumn{1}{c|}{ (2) }  &
\multicolumn{1}{c|}{ (3) }  &
\multicolumn{1}{c|}{ (4) }  &
\multicolumn{1}{c|}{ (5) }  &
\multicolumn{1}{c|}{ (6) }  &
\multicolumn{1}{c|}{ (7) }  &
\multicolumn{1}{c|}{ (8) }  &
\multicolumn{1}{c|}{ (9) }  &
\multicolumn{1}{c|}{ (10) } &
\multicolumn{1}{c|}{ (11) } &  \\
\\[-0.2cm] \hline \\[-0.2cm]
 \qq& HIPASSJ0626+24 & 9.9& 18.5&24.7& 13.7& 1.5&17.9& 20.6&17.6& 13.7 x 8.1 &1.5 x 0.9 \\ 
 \qq& UGC3600        &26.9& 75.9&24.0& 40.0& 1.8&16.1& 74.4&15.7& 40.0 x 10.8&1.8 x 0.5 \\
\end{tabular}
 }%
 }%
 \end{center} 
  \end{table*}
  
 \newpage 
\begin{table*}
\setcaptionmargin{0mm} \onelinecaptionstrue
\captionstyle{flushleft}
\renewcommand{\baselinestretch}{1.0}
\begin{center}
\rotatebox{-90}{%
\parbox{\textwidth}{%
{\bf Table 2.} Photometric paramerets I. (Condt.) \\
\vspace{0.3cm}
\begin{tabular}{r|l|r|r|r|r|r|r|r|r|c|c|c} \hline \\[-0.2cm]
\multicolumn{1}{c|}{} &
\multicolumn{1}{c|}{Name} &
\multicolumn{1}{c|}{$R_{\rm 50}(g)$}   &
\multicolumn{1}{c|}{$R_{\rm 90}(g)$}  &
\multicolumn{1}{c|}{$\mu_{\rm eff}(g)$} &
\multicolumn{1}{c|}{$R^{25}(g)$} &
\multicolumn{1}{c|}{$a_{\rm 25}$} &
\multicolumn{1}{c|}{$g^{25}$} &
\multicolumn{1}{c|}{$R^{26.5}(g)$} &
\multicolumn{1}{c|}{$g^{26.5}$} &
\multicolumn{1}{c|}{Opt.size} &
\multicolumn{1}{c|}{Opt.size} \\
&
\multicolumn{1}{c|}{}  &
\multicolumn{1}{c|}{$''$}  &
\multicolumn{1}{c|}{$''$}  &
\multicolumn{1}{c|}{$^{m}/\square''$}  &
\multicolumn{1}{c|}{$''$}  &
\multicolumn{1}{c|}{\tiny kpc}  &
\multicolumn{1}{c|}{\tiny mag}  &
\multicolumn{1}{c|}{$''$}  &
\multicolumn{1}{c|}{\tiny mag}  &
\multicolumn{1}{c|}{$''$}  &
\multicolumn{1}{c|}{\tiny kpc}  \\
&
\multicolumn{1}{c|}{ (1) }  &
\multicolumn{1}{c|}{ (2) }  &
\multicolumn{1}{c|}{ (3) }  &
\multicolumn{1}{c|}{ (4) }  &
\multicolumn{1}{c|}{ (5) }  &
\multicolumn{1}{c|}{ (6) }  &
\multicolumn{1}{c|}{ (7) }  &
\multicolumn{1}{c|}{ (8) }  &
\multicolumn{1}{c|}{ (9) }  &
\multicolumn{1}{c|}{ (10) } &
\multicolumn{1}{c|}{ (11) } & \\
 \\[-0.2cm] \hline \\[-0.2cm]
 \qq& J0929+2502     & 5.5&11.0&25.3&  5.3&0.7&19.7& 10.0&19.1&  5.3 x 4.7 &0.7 x 0.6 \\ 
 \qq& J0929+1155     &12.6&30.6&24.1& 19.5&2.3&17.3& 31.4&17.0& 19.5 x 7.8 &2.3 x 0.9 \\
\end{tabular}
 }%
 }%
 \end{center}
  \end{table*}
\clearpage

\clearpage
\headsep 1cm
\renewcommand{\baselinestretch}{1.0}

\setcounter{qub}{0}
\hoffset=0cm
\begin{table*}
\setcaptionmargin{0mm} \onelinecaptionstrue
\captionstyle{flushleft}
                                                               
\begin{center}
\rotatebox{-90}{%
\parbox{\textwidth}{%
{\bf Table 3.} Photometric paramerets II. \\
\vspace{0.3cm}
\begin{tabular}{r|l|r|r|r|r|c|c|c|} \hline \\[-0.2cm]
\multicolumn{1}{c|}{} &
\multicolumn{1}{c|}{Name} &
\multicolumn{1}{c|}{$g_{\rm tot}$}   &
\multicolumn{1}{c|}{$(u-g)_{\rm tot}$}  &
\multicolumn{1}{c|}{$(g-r)_{\rm tot}$} &
\multicolumn{1}{c|}{$(r-i)_{\rm tot}$} &
\multicolumn{1}{c|}{$(u-g)_{\rm per}$} &
\multicolumn{1}{c|}{$(g-r)_{\rm per}$} &
\multicolumn{1}{c|}{$(r-i)_{\rm per}$} \\
&
\multicolumn{1}{c|}{}  &
\multicolumn{1}{c|}{mag}  &
\multicolumn{1}{c|}{mag}  &
\multicolumn{1}{c|}{mag}  &
\multicolumn{1}{c|}{mag}  &
\multicolumn{1}{c|}{mag}  &
\multicolumn{1}{c|}{mag}  &
\multicolumn{1}{c|}{mag}  \\
&
\multicolumn{1}{c|}{ (1) }  &
\multicolumn{1}{c|}{ (2) }  &
\multicolumn{1}{c|}{ (3) }  &
\multicolumn{1}{c|}{ (4) }  &
\multicolumn{1}{c|}{ (5) }  &
\multicolumn{1}{c|}{ (6) }  &
\multicolumn{1}{c|}{ (7) }  &
\multicolumn{1}{c|}{ (8) }  \\
\\[-0.2cm] \hline \\[-0.2cm]
 \qq& HIPASSJ0626+24 &17.53$\pm$0.02&0.88$\pm$0.03&0.29$\pm$0.01&0.15$\pm$0.01&0.93$\pm$0.06&0.31$\pm$0.02&0.16$\pm$0.02\\
 \qq& UGC3600 &15.57$\pm$0.01&1.01$\pm$0.01&0.29$\pm$0.01&0.12$\pm$0.01&1.09$\pm$0.02&0.29$\pm$0.01&0.15$\pm$0.01\\
\end{tabular}
 }%
 }%
 \end{center}
  \end{table*}

  \newpage
\begin{table*}
\setcaptionmargin{0mm} \onelinecaptionstrue
\captionstyle{flushleft}
\begin{center}
\rotatebox{-90}{%
\parbox{\textwidth}{%
{\bf Table 3.} Photometric paramerets II. (Condt.) \\
\vspace{0.3cm}
\begin{tabular}{r|l|r|r|r|r|c|c|c|} \hline \\[-0.2cm]
\multicolumn{1}{c|}{} &
\multicolumn{1}{c|}{Name} &
\multicolumn{1}{c|}{$g_{\rm tot}$}   &
\multicolumn{1}{c|}{$(u-g)_{\rm tot}$}  &
\multicolumn{1}{c|}{$(g-r)_{\rm tot}$} &
\multicolumn{1}{c|}{$(r-i)_{\rm tot}$} &
\multicolumn{1}{c|}{$(u-g)_{\rm per}$} &
\multicolumn{1}{c|}{$(g-r)_{\rm per}$} &
\multicolumn{1}{c|}{$(r-i)_{\rm per}$} \\
&
\multicolumn{1}{c|}{}  &
\multicolumn{1}{c|}{mag}  &
\multicolumn{1}{c|}{mag}  &
\multicolumn{1}{c|}{mag}  &
\multicolumn{1}{c|}{mag}  &
\multicolumn{1}{c|}{mag}  &
\multicolumn{1}{c|}{mag}  &
\multicolumn{1}{c|}{mag}  \\
&
\multicolumn{1}{c|}{ (1) }  &
\multicolumn{1}{c|}{ (2) }  &
\multicolumn{1}{c|}{ (3) }  &
\multicolumn{1}{c|}{ (4) }  &
\multicolumn{1}{c|}{ (5) }  &
\multicolumn{1}{c|}{ (6) }  &
\multicolumn{1}{c|}{ (7) }  &
\multicolumn{1}{c|}{ (8) }  \\
 \\[-0.2cm] \hline \\[-0.2cm]
  \qq& J0929+2502 &18.95$\pm$0.02&0.87$\pm$0.05&0.16$\pm$0.03&0.11$\pm$0.04&0.94$\pm$0.08&0.19$\pm$0.05&0.13$\pm$0.06\\
 \qq& J0929+1155 &16.89$\pm$0.01&0.95$\pm$0.02&0.22$\pm$0.01&0.15$\pm$0.01&1.04$\pm$0.08&0.35$\pm$0.04&0.20$\pm$0.04\\
\end{tabular}
 }%
 }%
 \end{center}
  \end{table*}
  \clearpage

\clearpage
\headsep 1cm
\renewcommand{\baselinestretch}{1.0}

\setcounter{qub}{0}
\hoffset=0cm
\begin{table*}
\setcaptionmargin{0mm} \onelinecaptionstrue
\captionstyle{flushleft}
                                                               
\begin{center}
\rotatebox{-90}{%
\parbox{\textwidth}{%
{\bf Table 4.} Photometric paramerets III. \\
\vspace{0.3cm}
\begin{tabular}{r|l|r|r|r|r|r|r|} \hline \\[-0.2cm]
\multicolumn{1}{c|}{} &
\multicolumn{1}{c|}{Name} &
\multicolumn{1}{c|}{$\mu_{\rm 0}(g)$}   &
\multicolumn{1}{c|}{$\mu_{\rm 0}(r)$}  &
\multicolumn{1}{c|}{$\mu_{\rm 0}(B)$}  &
\multicolumn{1}{c|}{$\mu_{\rm 0,c,i}(B)$}  &
\multicolumn{1}{c|}{${\alpha }(g)$} &
\multicolumn{1}{c|}{$n(g)$} \\
&
\multicolumn{1}{c|}{}  &
\multicolumn{1}{c|}{$^{m}/\square''$}  &
\multicolumn{1}{c|}{$^{m}/\square''$}  &
\multicolumn{1}{c|}{$^{m}/\square''$}  &
\multicolumn{1}{c|}{$^{m}/\square''$}  &
\multicolumn{1}{c|}{\arcsec}  &
\multicolumn{1}{c|}{\tiny Sersic}  \\
&
\multicolumn{1}{c|}{ (1) }  &
\multicolumn{1}{c|}{ (2) }  &
\multicolumn{1}{c|}{ (3) }  &
\multicolumn{1}{c|}{ (4) }  &
\multicolumn{1}{c|}{ (5) }  &
\multicolumn{1}{c|}{ (6) }  &
\multicolumn{1}{c|}{ (7) }  \\
\\[-0.2cm] \hline \\[-0.2cm]
 \qq& HIPASSJ0626+24 &23.54$\pm$0.06 & 22.79$\pm$0.05 & 24.00 & 22.83 & 8.68$\pm$0.38 & 1.64$\pm$0.10\\ 
 \qq& UGC3600 &22.92$\pm$0.08 & 22.55$\pm$0.07 & 23.26 & 24.31 & 11.57$\pm$0.35 & 1.00$\pm$0.01\\
\end{tabular}
 }%
 }%
 \end{center}
  \end{table*}

\newpage
\begin{table*}
\setcaptionmargin{0mm} \onelinecaptionstrue
\captionstyle{flushleft}
\begin{center}
\rotatebox{-90}{%
\parbox{\textwidth}{%
{\bf Table 4.} Photometric paramerets III. (Condt.) \\
\vspace{0.3cm}
\begin{tabular}{r|l|r|r|r|r|r|r|} \hline \\[-0.2cm]
\multicolumn{1}{c|}{} &
\multicolumn{1}{c|}{Name} &
\multicolumn{1}{c|}{$\mu_{\rm 0}(g)$}   &
\multicolumn{1}{c|}{$\mu_{\rm 0}(r)$}  &
\multicolumn{1}{c|}{$\mu_{\rm 0}(B)$}  &
\multicolumn{1}{c|}{$\mu_{\rm 0,c,i}(B)$}  &
\multicolumn{1}{c|}{${\alpha }(g)$} &
\multicolumn{1}{c|}{$n(g)$} \\
&
\multicolumn{1}{c|}{}  &
\multicolumn{1}{c|}{$^{m}/\square''$}  &
\multicolumn{1}{c|}{$^{m}/\square''$}  &
\multicolumn{1}{c|}{$^{m}/\square''$}  &
\multicolumn{1}{c|}{$^{m}/\square''$}  &
\multicolumn{1}{c|}{\arcsec}  &
\multicolumn{1}{c|}{\tiny Sersic}  \\
&
\multicolumn{1}{c|}{ (1) }  &
\multicolumn{1}{c|}{ (2) }  &
\multicolumn{1}{c|}{ (3) }  &
\multicolumn{1}{c|}{ (4) }  &
\multicolumn{1}{c|}{ (5) }  &
\multicolumn{1}{c|}{ (6) }  &
\multicolumn{1}{c|}{ (7) }  \\
 \\[-0.2cm] \hline \\[-0.2cm]
 \qq& J0929+1155 &22.88$\pm$0.35 & 22.77$\pm$0.10 & 23.14 & 24.15 & 6.73$\pm$1.73 & 1.12$\pm$0.19\\ 
 \qq& J0931+2717 &22.58$\pm$0.47 & 22.81$\pm$0.43 & 22.74 & 23.75 & 3.91$\pm$1.27 & 1.07$\pm$0.19\\
\end{tabular}
 }%
 }%
 \end{center}
  \end{table*}

\clearpage

}


\begin{thebibliography}{99}

\bibitem{joeveer78}
\refitem{article}
 M. J\"oeveer, J. Einasto, and E. Tago, \mnras\ {\bf 185}, 357 (1978).

\bibitem{kirshner81}
\refitem{article}
 R.~P. Kirshner, A.~Oemler~Jr., P. L. Schechter, and S.~A. Shectman, 
\apj\ {\bf 248}, L57 (1981).

\bibitem{montero09}
\refitem{article}
 A.~D. Montero-Dorta and F. Prada, \mnras\ {\bf 399}, 1106 (2009).

\bibitem{deLapparent95}
\refitem{article}
 V. de~Lapparent,
in {\it Proc. Les~Houches Summer School, Session~LX, Cosmology and
Large Scale Structure}, Ed. by R. Schaeffer, J. Silk, M. Spiro,
and J. Zinn-Justin (Elsevier Sci. Publ. Co., Amsterdam, 1996),
p.~107.

\bibitem{Peebles01}
\refitem{article}
 P.~J.~E.~Peebles, \apj\ {\bf 557}, 459 (2001).

\bibitem{Tikhonov09}
\refitem{article}
 A.~V. Tikhonov and A.~A. Klypin, \mnras\ {\bf 395}, 1915 (2009).

\bibitem{Rojas04}
\refitem{article}
 R.~R. Rojas, M.~S. Vogeley, F. Hoyle, and J.~Brinkmann, \apj\ {\bf 617}, 50 (2004).

\bibitem{Rojas05}
\refitem{article}
 R.~R. Rojas, M.~S. Vogeley, F. Hoyle, and J.~Brinkmann, \apj\ {\bf 624}, 571 (2005).

\bibitem{patiri06}
\refitem{article}
 S.~G. Patiri, F. Prada, J. Holtzman, et al., \mnras\ {\bf 372}, 1710 (2006).

\bibitem{Hoyle12}
\refitem{article}
 F. Hoyle, M.~S. Vogeley, and D. Pan, \mnras\ {\bf 426}, 3041 (2012).

\bibitem{kreckel11}
\refitem{article}
 K. Kreckel, M.~R. Joung, and R. Cen, \apj\ {\bf 735}, 132 (2011).

\bibitem{Einasto06}
\refitem{article}
 J.~E. Einasto,
 Colloquim on Cosmic Voids,
 {\tt http://www.astro.rug.nl/$\sim$weygaert/\\/knawvoid.program.php}


\bibitem{cross02}
\refitem{article}
 N. Cross and S.~P. Driver, \mnras\ {\bf 329}, 579 (2002).

\bibitem{mihos97}
\refitem{article}
 J.~C. Mihos, S.~S. McGaugh, and W.~J.~G. de Block, \apj\ {\bf 477}, L79 (1997).

\bibitem{PaperI}
\refitem{article}
 S.~A. Pustilnik and A.~L. Tepliakova, \mnras\ {\bf 415}, 1188 (2011).

\bibitem{PaperII}
\refitem{article}
 S.~A. Pustilnik, A.~L. Tepliakova, and A.~Y. Kniazev, \ab\  {\bf 66}, 255 (2011).

\bibitem{PaperIII}
\refitem{article}
 S.~A. Pustilnik, J.-M. Martin, A.~L. Tepliakova, and A.~Y. Kniazev, \mnras\ {\bf 417}, 1335 (2011).

\bibitem{Triplet}
\refitem{article}
 J.~N. Chengalur and S.~A. Pustilnik, \mnras\ {\bf 428}, 1579 (2013).

\bibitem{kreckel12}
\refitem{article}
 K. Kreckel, E. Platen, M.~A. Aragon-Calvo, et al., \aj\ {\bf 144}, 16 (2012).

\bibitem{DR7}
\refitem{article}
 K.~N. Abazajian, J.~K. Adelman-McCarthy, M.~A.Ag\"ueros, et al., \apjs\ {\bf 182}, 543 (2009).

\bibitem{pegase}
\refitem{article}
 M. Fioc, B. Rocca-Volmerange,\\ arXiv:astro-ph/9912179 (1999).

\bibitem{Garcia09}
\refitem{article}
 D.~A. Garcia-Appadoo, A.~A. West, J.~J. Dalcanton, et al., \mnras\ {\bf 394}, 340 (2009).

\bibitem{hipass}
\refitem{article}
 M.~J. Meyer, M.~A. Zwaan, R.~L. Webster, et al., \mnras\ {\bf 350}, 1195 (2004).

\bibitem{haynes11}
\refitem{article}
 M.~P. Haynes, R. Giovanelli, A.~M. Martin, et al., \apj\ {\bf 142}, 170 (2011).

\bibitem{York}
\refitem{article}
 D.~G. York, J. Adelman, J.~E. Anderson, et al., \aj\ {\bf 120}, 1579 (2000).

\bibitem{LSB-SDSS}
\refitem{article}
 A.~Y. Kniazev, E.~K. Grebel, S.~A. Pustilnik, et al., \aj\ {\bf 127},
704 (2004).

\bibitem{Lupton05}
\refitem{article}
 R. Lupton,
 {\tt http://www.sdss.org/dr7/\\/algorithms/sdssUBVRITransform.html\#\\\#Lupton2005}

\bibitem{holmberg}
\refitem{article}
 E. Holmberg, Lund Medd. Astron. Obs. Ser. {\bf 136}, 1 (1958).

\bibitem{Sersic}
\refitem{article}
 J.~L. Sersic, Boletin de la Asociacion Argentina de Astronomia {\bf 6},
99 (1963).

\bibitem{SF11}
\refitem{article}
 E. F. Schlafly and D. P. Finkbeiner, \apj\ {\bf 737}, 103 (2011).

\bibitem{Matthews99}
\refitem{article}
 L. D. Matthews, J. S. Gallagher, W. van Driel, \aj\ {\bf 118}, 2751 (1999).

\bibitem{Kruit81}
\refitem{article}
 P.~C. van der Kruit and L. Searle,  \aaa\ {\bf 95}, 105 (1981).

\bibitem{Roych13}
\refitem{article}
 S. Roychowdhury, J.~N. Chengalur, I.~D.~Karachentsev, and E.~I. Kaisina,
\mnras\ {\bf 436}, 104 (2013).

\bibitem{Salpeter}
\refitem{article}
 E.~E. Salpeter, \apj\ {\bf 121}, 161 (1955).

\bibitem{kroupa}
\refitem{article}
 P. Kroupa, C.~A. Tout, and G. Gilmore, \mnras\ {\bf 262}, 545 (1993).

\bibitem{Tucker06}
\refitem{article}
 D.~L. Tucker, S. Kent, M.~W. Richmond, et al., Astronomische Nachrichten 
{\bf 327}, 821 (2006).

\bibitem{Roberts69}
\refitem{article}
 M.~S. Roberts, \aj\ {\bf 74},  859 (1969).

\bibitem{Zibetti09}
\refitem{article}
 S. Zibetti, S. Charlot, and H.-W. Rix, \mnras\ {\bf 400}, 1181 (2009).

\bibitem{Petro}
\refitem{article}
 V. Petrosian, \apj\ {\bf 209}, 1 (1976).

\bibitem{Pustilnik02}
\refitem{article}
 S.~A. Pustilnik, J.-M. Martin, W.~K. Huchtmeier, et al., \aaa\ {\bf 389},
405 (2002).

\bibitem{Pustilnik03}
\refitem{article}
 S.~A. Pustilnik, A.~Y. Kniazev, A.~G. Pramsky, et al., \aaa\ {\bf 409},
917 (2003).

\bibitem{DDO68}
\refitem{article}
 S.~A. Pustilnik, A.~Y. Kniazev, and A.~G. Pramsky, \aaa\ {\bf 443}, 91
(2005).

\bibitem{DDO68_sdss}
\refitem{article}
 S.~A. Pustilnik, A.~L. Tepliakova, and A.~Y. Kniazev, Astronomy Letters
{\bf 34}, 457 (2008).

\bibitem{J0926}
\refitem{article}
 S.~A. Pustilnik, A.~L. Tepliakova, A.~Y. Kniazev, and A.~N. Burenkov, \mnras\ 
{\bf 401}, 333 (2010).

\bibitem{SBS0335BTA}
\refitem{article}
 S.~A. Pustilnik, A.~G. Pramskij, and  A.~Y. Kniazev, \aaa\ {\bf 425}, 51
(2004).

\bibitem{Tikhonov14}
\refitem{article}
 N.~A. Tikhonov, O.~A. Galazutdinova, and V.~S.~Lebedev, Astronomy Letters
{\bf 40}, 1 (2014).

\bibitem{ekta08}
\refitem{article}
 Ekta, J.~N. Chengalur, and S.~A. Pustilnik, \mnras\ {\bf 391}, 881 (2008).

\bibitem{Bolshev83}
\refitem{book}
 L.~N. Bolshev and N.~V. Smirnov, {\it Tables of Mathematical Statistics}
(Nauka, Moscow, 1983) [in Russian].

\bibitem{Pustilnik95}
\refitem{article}
 S.~A. Pustilnik, A.~V. Ugryumov, V.~A. Lipovetsky, et al., \apj\ {\bf 443}, 499 (1995).


\end{thebibliography}
\end{document}